\documentclass[prl,twocolumn,showpacs,groupedaddress]{revtex4}

\usepackage{graphicx}

\begin{document}

\title{\bf Multilevel Tunnelling Systems and Fractal Clustering \\ 
in the Low-Temperature Mixed Alkali-Silicate Glasses }

\author{ Giancarlo Jug$^a$\cite{J} and Maksym Paliienko }

\affiliation{
Dipartimento di Fisica e Matematica, Universit\`a dell'Insubria, Via
Valleggio 11, 22100 Como, Italy \\
$^a$INFN -- Sezione di Pavia (Italy) and IPCF -- Sezione di Roma (Italy) }

\date{\today }

\begin{abstract}
The thermal and dielectric anomalies of window-type glasses at low
temperatures ($T<$ 1 K) are rather successfully explained by the
two-level systems (2LS) standard tunneling model (STM). However, the
magnetic effects discovered in the multisilicate glasses in recent
times, magnetic effects in the organic glasses and also some older data 
from mixed (SiO$_2$)$_{1-x}$(K$_2$O)$_x$ and 
(SiO$_2$)$_{1-x}$(Na$_2$O)$_x$ glasses indicate the need for a 
suitable extension of the 2LS-STM. We show that -- not only for 
the magnetic effects, but already for the mixed glasses in the absence 
of a field -- the right extension of the 2LS STM is provided by the 
(anomalous) multilevel tunnelling systems (A-TS) proposed by one of us
for multicomponent amorphous solids. Though a secondary type of TS, 
different from the standard 2LS, was invoked long ago already, we clarify 
their physical origin and mathematical description and show that their 
contribution considerably improves the agreement with the experimental 
data. 
\end{abstract}


\maketitle

\section{I. INTRODUCTION}



Glasses are ubiquitous materials of considerable importance for many practical 
applications, however for physicists the nature of the glass transition and 
the ultimate microscopic structure of glasses determining their physical 
properties remain to this day issues of considerable intellectual challenge 
\cite{Ref1}. Glasses are normally regarded as fully homogeneously disordered 
amorphous systems, much alike liquids except for the glassy arrested dynamics 
close and below the glass transition temperature $T_g$, which leads to an 
increase of several orders of magnitude in the viscosity for $T\to T_g^+$. 
Nevertheless this homogeneity is most probably only a useful idealization, for 
real glasses must always contain some small (in ceramic glasses not so small) 
concentration of tiny, ordered or nearly-ordered regions of variable size with 
their own frozen dynamics. Indeed the thermodynamically stable phase of an 
undercooled liquid would be the perfect crystal, thus every substance in 
approaching the crystallization temperature $T_c$ ($T_c>T_g$) from above would 
spontaneously generate local regions of enhanced regularity (RER) much like a 
system (a vapour or a paramagnet) approaching its critical temperature is 
known to develop regions (droplets) resembling the ordered low-temperature 
phase. These RER are of course to be distinguished from the concept of 
short-ranged atomic order which is typical of ideal glasses and is restricted 
to the first few atomic spacings. We are considering in this paper realistic 
glasses in which a degree of {\em devitrification} has occurred. The size and 
concentration of these RER will depend, e.g., on the rapidity of the quench 
leading to the formation of the glass, but also on the chemical composition of 
the substance, the presence of impurities and so on. However, on general 
grounds, even the purest of glasses should contain RER in non-zero 
concentration and size. 
 
That this is the case has been demonstrated recently for the structure of the 
metallic glass Zr$_{50}$Cu$_{45}$Al$_5$ \cite{Ref2}, where a combination of 
fluctuation electron spectroscopy (FEM) and Monte Carlo simulation (MC) has 
revealed the presence of crystalline regions of sub-nanometer size embedded in
an otherwise homogeneously amorphous mass of the same composition. It is 
believed that other metallic glasses should present similar structural features 
and thus -- on general grounds -- one would expect that non-metallic window 
glasses too, like pure SiO$_2$ and all the more so the commercial 
multisilicates of complex chemical composition, should present a multiphased 
structure with the size and concentration of the near-crystalline regions, or 
RER, depending (e.g.) on composition, quench rate and the presence of impurities 
acting as nucleation centres for the RER. Indeed, materials of the general 
composition (MgO)$_x$(Al$_2$O$_3$)$_y$(SiO$_2$)$_{1-x-y}$ (MAS, in short) are 
termed {\em ceramic glasses} (one of the best-known commercial examples being 
Schott's Ceran where Li$_2$O replaces MgO, and of course CaO or BaO can also 
replace or be added to MgO and still yield a ceramic glass). These materials 
are known to contain micro-crystals embedded in an otherwise homogeneously 
amorphous matrix \cite{Ref3}. This is not surprising, for materials made up of 
a good glass-former (e.g. SiO$_2$, Al$_2$O$_3$ ...) and good crystal-formers 
(e.g. BaO, K$_2$O ...) are known to be multiphased \cite{Ref4} with the good 
crystal-formers generating their own pockets and channels carved out within 
the otherwise homogeneously amorphous network of the good glass-former's ions 
\cite{Ref5}. Within these pockets and channels, incipient nano- or even 
micro-crystals may form, but the point of view will be taken in this work 
that on general grounds even the purest, single-component (e.g. As, SiO$_2$) 
glass-former will be rich in RER unless the quench-rate from the melt is so
large as to avoid the formation of crystalline regions or RER.       

These refined structural details of glasses are evidently hard to reveal in 
all and especially the near-ideal cases (no good crystal-formers, no 
impurities added, and rapid quenches) with the available spectroscopic 
techniques. For example X-ray spectroscopy does not reveal nano-crystals below 
the nanometer size. However, at low- and very low-temperatures -- where all 
said structural features remain basically unaltered -- some recent experimental 
findings might now improve perspectives with what would appear set to become a 
new spectroscopy tool. Indeed a series of remarkable magnetic effects have 
recently been discovered in non-magnetic glasses (multisilicates and organic 
glasses) \cite{Ref6,Ref6b,Ref7,Ref8,Ref9} with, in the opinion of the present 
Authors, a most likely explanation for the new phenomena stemming precisely 
from the multiphase nature of real glasses and the presence of the RER or 
micro-crystalline regions in their microscopic structure. In turn, when the 
multiphase theory shall be fully developed, the magnetic effects could 
represent a valid new spectroscopic tool capable of characterizing micro- or
nano-crystals or even incipient crystals and RER in the real glasses. The key 
to this possible development is some new exciting physics of the cold glasses 
in the presence (and even in the absence, as shown in the present paper) of a 
magnetic field. The magnetic effects in the cold glasses could become, 
eventually, the amorphous counterpart of the de Haas - van Alphen and 
Shubnikov - de Haas effects in crystalline solids in determining the real 
structure of amorphous solids.  
      
Systematic research in the low-temperature properties of glasses has been 
on-going for more than 40 years and some significant theoretical and 
experimental progress has been made in the understanding of the unusual 
behaviour of glasses and of their low-temperature anomalies \cite{Ref10}. 
This temperature range ($T<$1 K) is deemed important for the appearance of 
universal behaviour (independent of composition), as well as for the effects 
of quantum mechanics in the physics of glasses. However, to make progress in 
the understanding of the low-temperature physics of glasses there remains a 
wide range of important questions that are still open or only partially 
answered, particularly in the light of some still poorly understood recent, 
and even older, experiments in cold composite glasses. 
	
It is well known that cold glasses show somewhat universal thermal, acoustic 
and dielectric properties which are very different from those of crystalline 
solids at low temperatures \cite{Ref11}. Below 1 K, the heat capacity $C_p$ 
of dielectric glasses is much larger and the thermal conductivity $\kappa$ is 
orders of magnitude lower than the corresponding values found in their 
crystalline counterparts. $C_p$ depends approximately linearly and $\kappa$ 
almost quadratically on temperature $T$, whereas in crystals one can observe a 
cubic dependence for both properties, well understood in terms of the 
Debye-Einstein's theory of lattice vibrations. The dielectric constant (real 
part) $\epsilon'$ and sound velocity at low frequencies display in glasses a 
universal logarithmic dependence in $T$. These ``anomalous'' and yet universal 
thermal, dielectric and acoustic properties of glasses are well explained (at 
least for $T<$ 1 K) since 1972 when Phillips \cite{Ref12} and also Anderson, 
Halperin and Varma \cite{Ref13}, independently, introduced the tunnelling model 
(TM), the fundamental postulate of which was the general existence of atoms or 
small groups of atoms in cold amorphous solids which can tunnel like a single 
quantum-mechanical particle between two configurations of very similar energy 
(two-level systems, 2LS). The 2LS TM is widely used in the investigation of 
the low-temperature properties of glasses, mostly because of its technical 
simplicity. In fact, it will be argued in this paper that tunneling takes 
place in more complicated local potential scenarios (multiwelled potentials) 
and a situation will be discussed where the use of a number of ``states'' 
greater than two is essential. Moreover, new insight will be given on the role 
of percolation and fractal theory in the TM of multicomponent glasses. We 
present in this paper the justification and details of the construction of an 
extended TM that has been successfully employed to explain the unusual 
properties of the cold glasses in a magnetic field \cite{Ref14}, as well as in 
zero field when systematic changes in the glass' composition are involved 
\cite{Ref15}.  

The linear dependence in $\ln(T)$ of the real-part of the dielectric constant 
$\epsilon'(T)$ makes the cold glasses useful in low-temperature thermometry 
and, normally, structural window-type glasses are expected to be isotropic 
insulators that do not present any remarkable magnetic-field response 
phenomena (other than a weak response in $C_p$ to the trace paramagnetic 
impurities). For some multicomponent silicate glass it has become possible to 
measure observable, much larger than expected changes in $\epsilon'(T,B)$ 
($\delta\epsilon'/\epsilon'\sim 10^{-4})$ already in a magnetic field as weak 
as a few Oe \cite{Ref6}. A typical glass giving such strong response has the 
composition Al$_2$O$_3$-BaO-SiO$_2$: thus a MAS ceramic-glass, herewith 
termed AlBaSiO. The measurements were made on thick sol-gel fabricated films, 
a fabrication procedure favoring micro-crystal formation \cite{Ref4}, cooled 
in a $^3$He-$^4$He dilution refrigerator reaching temperatures as low as 6 mK. 
Magnetic effects have been reported for both the real- and imaginary-part of 
$\epsilon$ at low frequency ($\omega\sim$ 1 kHz), for the heat capacity $C_p$
(see e.g. \cite{Ref14}) and for the polarization echo (where changes in the 
presence of a magnetic field have been the strongest \cite{Ref8}) as well. 
This behavior was confirmed in other multicomponent glasses, like borosilicate 
optical glass BK7 and commercial Duran \cite{Ref6b} and, moreover, similar 
effects on $\epsilon'(T)$ have been confirmed in studies of the structural 
glass $a$-SiO$_{2+x}$C$_y$H$_z$ in the range 50$<T<$400mK and $B\leq$3T 
\cite{Ref7}. Although the dielectric magnetocapacitance enhancement is not 
dramatic ($\delta\epsilon'(B)/\epsilon'$ is typically in the 10$^{-6}$ - 
10$^{-4}$ range), the available measurements show that an unusual effect 
of the magnetic field is indeed present in the above glasses, yet not 
measurable in ultra-pure SiO$_2$ (Suprasil W), and cannot be ascribed to 
spurious agents \cite{Ref16} or to trace paramagnetic impurities (always 
present in silicate glasses, although in $<$6 ppm concentration in the case 
of BK7). Polarization-echo experiments in the AlBaSiO, Duran and BK7 glasses 
have also shown considerable sensitivity in the response of the echo amplitude 
to very weak magnetic fields and the magnetic effects clearly do not scale 
with the concentration of paramagnetic impurities \cite{Ref6b,Ref8}. Striking 
magnetic effects, the presence of a novel isotope effect and remarkable 
oscillations in the dephasing time have also been reported in studies of the 
polarization echos in organic glasses (amorphous glycerol) \cite{Ref9}. 
However, in terms of a detailed theoretical justification for all of the 
observed magnetic effects (and the lack of an observable magnetic effect in 
the acoustic response \cite{Ref17}, so far) an explanation relying on a single 
theoretical model for all of the available experimental data is still missing. 
We believe the two-phase model reproposed in this paper to be the correct 
generalization of the standard 2LS TM that is being sought and here we 
work out its predictions in zero magnetic field, but for different controlled 
concentrations of glass-forming and crystal-forming components. In this way, 
we put our approach to a new test.

The essential behavior of the dielectric response of glasses at low 
temperatures is well known \cite{Ref11}: starting from the lowest temperatures, 
the dielectric constant $\epsilon'$ first decreases with increasing 
temperature due to the resonant interaction of the electric field with the 
tunneling systems (TS). According to the standard 2LS TM (STM from now on), 
the dielectric constant is predicted to vary like -$\ln T$ due to the constant 
density of states of the TS. Above a certain temperature $T_0(\omega)$ 
relaxational absorption of the TS becomes important, resulting in an increase 
of the dielectric constant with temperature proportional to +$\ln T$ according 
to the STM. This has been checked experimentally for several glasses. The 
temperature $T_0$ of the resulting minimum depends on the frequency $\omega$ 
and occurs around 50 to 100 mK in measurements at about 1 kHz: in AlBaSiO 
glass, the position of the minimum has been shown to shift considerably when 
the amplitude of the driving electric field is considerably increased 
(probably, the consequence of non-linearity setting in).

Some more  interesting behavior has been shown by some as yet unexplained 
data from experiments on the mixed 
(SiO$_2$)$_{1-x}$(K$_2$O)$_x$ and (SiO$_2$)$_{1-x}$(Na$_2$O)$_x$ 
glasses, studied as a function of the concentration $x$ of the good 
crystal-former at low temperatures \cite{Ref18}. The heat capacity $C_p(T)$ for 
these glasses is larger than that for pure vitreous silica and the behavior as 
a function of $T$ is very peculiar for different molar concentrations $x$ of 
potassium or sodium oxide and is not explained by the STM. The heat capacity 
decreases and then increases again with increasing molar concentration $x$ of 
K$_2$O. The minimum in the dielectric constant $\epsilon'(T)$ is observed for 
$T_0$ near 100 mK as is typical for these glassy solids. The temperature 
dependence of $\epsilon'$, both above and below $T_0$, shows however a slope 
in $\pm\ln T$ qualitatively increasing with increasing concentration $x$ of 
K$_2$O. One can notice, moreover, that above the minimum $T_0$ the relaxation 
part of $\epsilon'$ is increasing faster in slope than the resonant part below
$T_0$ for the same $x$ \cite{Ref18}, a feature completely unexplained thus far. 
This work is an indication that not only the magnetic and electric fields 
influence the properties of glasses, but the concentration of chemical species
in the composite materials too (a fact not accounted for by the STM). In this 
paper we show in detail how the very same approach that explains the magnetic 
properties in the multisilicates \cite{Ref14} also provides a quantitative 
explanation for the above-mentioned composition-dependent physical properties. 
The picture that emerges regarding the nature of the TS in the multicomponent 
glasses provides a novel and detailed description of the micro- and 
nano-structure of the glassy state. In turn, the linear dependence of the 
concentration $x_{ATS}$ of anomalous TS (ATS) -- those responsible for the 
magnetic and composition effects in our theory -- on $x$ fully corroborates 
the founding assumptions of our approach.

The paper is organised as follows. In Section II we present a detailed 
justification for the two-phase approach and the construction of the 
two-species TS model for the amorphous solids at low temperatures. In Section 
III we present the detailed predictions of this model for the dielectric 
constant $\epsilon'(T,x)$ as a function of temperature $T$ and composition 
$x$ of alkali oxide (good-crystal former) for the mixed glasses and we compare
the predictions with the experimental data \cite{Ref18}. In Section IV we 
present the detailed predictions of our model for the heat capacity $C_p(T,x)$
for the mixed glasses and we compare the predictions with the available 
experimental data \cite{Ref18}. Section V contains our conclusions about the 
nature of the TS, namely we show how the tunneling ``particle'' must in fact 
represent a whole cluster of $N$ correlated real tunneling ions in the 
material. Finally in the Appendix we work out how the effective tunneling 
parameters of our model are related, via $N$, to more standard microscopic 
tunneling parameters. A short preliminary account of this work was published 
in \cite{Ref15}.

\section{II. BUILDING UP A SUITABLE TUNNELING MODEL}
   
The traditional picture \cite{Ref11} viewed the TS, present in low 
concentration ($\sim 10^{16}$ g$^{-1}$) in the material, associated with the 
non-equivalence of two (or more) bonding-angle configurations per atomic unit 
in the amorphous solid's atomic structure. Each TS is represented in the 
standard case by a particle in an asymmetric (one-dimensional (1D)) 
double-well potential where, at low $T$, only the ground states of the two 
constituent single wells are assumed to be relevant. Consequently, only the 
two lowest-lying double-well states are taken to determine the physics of 
each single TS. A 2LS simplified picture then applies and one can describe 
the low energy Hamiltonian of each independent TS in terms of an equivalent 
notation with spin-$\frac{1}{2}$ pseudospin matrices $\sigma_x$ and 
$\sigma_z$ (Pauli matrices), leading to the compact notation 
$H_0^{(2)}=-\frac{1}{2}(\Delta\sigma_z+\Delta_0\sigma_x)$ for the Hamiltonian 
of a single 2LS TS. In matrix form (the so-called well- or position-space 
representation, $\langle i|H_0^{(2)}|j\rangle$, $|i\rangle$ being the two 
unequivalent wells, $i=1,2$ or $i=L, R$) this then reads:

\begin{equation}H_0^{(2)}=-\frac{1}{2} \left(\begin{array}{cc}
\Delta & \Delta_0 \\ \Delta_0 & -\Delta
\end{array}\right).
\label{rtH2matrix}  
\end{equation}
Here, the phenomenological parameters $\Delta$ and $\Delta_0$ (known as 
the energy asymmetry and (twice) the tunnelling matrix element, respectively) 
represent a way of describing the essential low-$T$ relevant features of the 
full, and yet unknown in its details, TS single-particle Hamiltonian in the 
effective single-well matrix representation. One obtains 
${\cal E}_{1,2}=\pm\frac{1}{2}\sqrt{\Delta^2+\Delta_0^2}$ for the two 
lowest-lying energy levels and the physics of the glass is then extracted by 
assuming (initially) the 2LS to be independent entities in the glass and 
averaging physical quantities over a probability distribution for the 
parameters $\Delta$, $\Delta_0$ of the standard form ($\bar{P}$ being a 
material-dependent constant):

\begin{equation}
P(\Delta,\Delta_0)=\frac{\bar{P}}{\Delta_0}.
 \label{rtdistrib}
\end{equation}
This distribution reflects the generally accepted opinion that $\Delta$ and 
$-\ln(\Delta_0/\hbar\Omega)$ (the latter proportional to the double-well 
potential barrier $V_0$ divided by the single-well attempt frequency $\Omega$, 
$V_0/\hbar\Omega$) should be rather broadly distributed in a homogeneously
disordered solid. This leads to an almost constant density of states (DOS) 
and the above STM has been employed with considerable success in order to 
explain a wide range of physical properties (thermal, dielectric (ac and 
pulsed), acoustic, and so on \cite{Ref11}) of non-metallic glasses below 1 K. 

There are, however, several drawbacks with the STM as thoughtfully pointed 
out by Leggett and co-workers \cite{Ref19}. For a start, the nature of the TS 
(and of the two wells of a single 2LS) and that of the motion inside a single 
TS remain to date completely unknown \cite{Ref20}. Much easier is the 
diagnostic for the nature of 2LS in the case of disordered crystals, such as 
Li-KCl or KBr-KCN solutions \cite{Ref23} (we shall come back to disordered 
crystals later). On general grounds, other types of (multilevel) excitations
are always possible in glasses and it is not clear why their distribution of 
parameters should be so similar (and given by Eq. (\ref{rtdistrib})) in all of 
the amorphous solids. Next, the STM has gathered great consensus for the 
explanation of many experiments at low temperatures, but in its simplest 
form (Eqs. (\ref{rtH2matrix})-(\ref{rtdistrib})) it fails to explain sound 
velocity shift and adsorption data at low-$T$ and the origin of the ``bump'' 
in $C_p$ (and ``plateau'' in $\kappa$) well above $T_0$ that goes under the 
name of {\em boson peak} (see e.g. the references in \cite{Ref19}). Moreover, 
the STM fails to explain the remarkable universality of the ultrasonic 
attenuation coefficient $Q^{-1}$ (roughly, independent of every external 
parameter and glass chemical composition) below 1 K \cite{Ref24}. To resolve 
these (and other) difficulties with the STM, Leggett and collaborators have 
proposed a generic model in the context of anharmonic elasticity theory which 
can account for all of the significant features of glasses below 1 K, 
including the super universality of $Q^{-1}$ \cite{Ref19}.    

However, it is hard to see how this generic elastic model can be extended to
account for the magnetic and composition-dependent effects in glasses, also
considering that in the multicomponent (i.e. real, not model) glasses most of 
the said universality features (e.g. in $C_p(T,B)$ and $\epsilon'(T,B)$ 
\cite{Ref6,Ref14} or in $C_p(T,x)$ and $\epsilon'(T,x)$ \cite{Ref15,Ref18}) 
are lost. Therefore, here we adopt the strategy of resuming the TS approach by 
means of a completely different (and more modern) justification for the TM, 
and then extend the STM to take  the presence of a magnetic field into account 
and to explain composition-dependent features (this work). In a rather general 
fashion, the TS can be thought of as arising from the shape of the theoretical 
potential-energy landscape 
$E(\{ {\bf r}_i\})$ of a glass as $T$ is lowered well below the glass freezing 
transition $T_g$. The concept of free-energy landscape was introduced, e.g., 
by Stillinger \cite{Ref25} and successfully employed in the study of glasses 
(e.g. \cite{Ref1}) and spin-glasses (e.g. \cite{Ref26}). A large number of 
local and global minima develop in $E(\{{\bf r}_i \})$ as $T\to 0$, the 
lowest-energy minima of interest being made up of $n_w=2, 3, \dots$ local 
wells separated by shallow energy barriers. At low-$T$ these 
configuration-space local multiwelled 
potentials are our TS and it seems reasonable to expect that the $n_w=2$ - 
welled potentials (2LS) should be ubiquitous in this picture. These should be 
thought of as an effective representation of local ``tremblements'' of the 
equilibrium positions $\{ {\bf r}_i^{(0)} \}$ of some of the glass atoms/ions' 
positions spanning over a large number of near-neighbors' distances (unlike 
in the case of disordered crystals, where the TS are known to be rather 
well-localized dynamical entities). Hence, just as the $n_w=2$ - welled case 
is possible, so ought to be the $n_w=3, 4, \dots$ - welled situations which 
would also be local rearrangements involving several atoms/ions/molecules. 
The concentration of these local potentials should not necessarily decrease 
exponentially with increasing $n_w$, in glasses, as it is known to happen for 
the disordered crystals (2LS present with probability $c^2$, 3LS with $c^3$, 
4LS with $c^4$ ... and so on, $c$ being the defects' percent concentration).

\begin{figure}[h]
\includegraphics[width=0.48\textwidth]{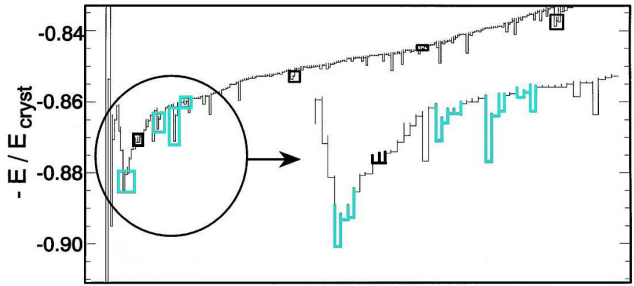} \vskip-3mm
\caption[2]{ (color online) The energy landscape (for $\rho$ =1 Lennard-Jones 
density, adapted from Ref. \cite{Ref27}) of a toy glass model, with 
highlighted multiwelled potentials (black the 2LS, light blue the 3LS, 4LS, 
...). }
\label{elandscape}
\end{figure}

We can reason this out over the quantitative description of the glassy energy 
landscape of a model situation, as was studied by Heuer \cite{Ref27} who 
considered the molecular-dynamics (MD) simulation data of a toy glass made up 
of several (13 or 32) particles interacting through a Lennard-Jones potential 
and with periodic boundary conditions applied. Adopting a suitable 1D 
projection procedure, where a ``distance'' between two local total energy 
minima is (not completely unambiguously) defined, the 1D position of a local 
minimum is somehow attained and the energy landscape of the model system can 
be charted out. Fig. \ref{elandscape} reports this chart for the total energy 
landscape for a 
given density (from \cite{Ref27}). Beside the deep minimum of the crystalline 
configuration, a large number of local minima is visualized and then a suitable 
definition of local double-welled potentials (2LS) is adopted to classify 
couples of adjacent minima constituting a single tunneling 2LS (highlighted in 
black, in Fig. \ref{elandscape}). This definition guarantees that at low 
temperatures a 
``particle'' subjected to any such local potentials will switch between both 
minima without escaping to a third minimum. Interestingly, the distribution of 
the tunneling parameters $\Delta, \Delta_0$ (suitably defined) for these 2LS 
could also be evaluated from MD simulations of the above toy model, and this
$P(\Delta,\Delta_0)$ turned out to be not so perfectly flat as a function of 
$\Delta$ as implied by Eq. (\ref{rtdistrib}). Rather, an increase (though no 
divergence) of probability for 2LS with $\Delta\to 0$ was measured in previous 
MD simulations \cite{Ref28}. Still, Fig. \ref{elandscape} also allows for 
tunneling multiwelled local potentials to be identified, and we have 
highlighted (in light blue) 
some of them (three- and four-welled local potentials). The requirement 
that a ``particle'' subjected to such multiwelled local potentials should not 
escape (at low-$T$) to foreign minima has been equally respected and one can 
see that these multiwelled situations are not at all rare. We therefore 
believe that 3LS, 4LS and so on should also be considered in the TM. The 
reduced Hamiltonians (well- or position-representation) for these local 
multiwelled potentials can be easily written down, as generalizations of 
Eq.  (\ref{rtH2matrix}). For $n_w=3$ (3LS):

\begin{equation}H_0^{(3)}=\left(\begin{array}{ccc}
E_1 & D_0 & D_0 \\ D_0 & E_2 & D_0 \\ D_0 & D_0 & E_3
\end{array}\right)
\label{rtH3matrix}  
\end{equation}
where $E_1$, $E_2$, $E_3$ are random energy asymmetries between the wells 
chosen to satisfy ${\sum}_{i=1}^{3}{E_i}=0$ and taken from an appropriate 
probability distribution (see below), together with the tunneling parameter 
$D_0>0$ (see below). For $n_w=4$ (4LS):

\begin{equation}H_0^{(4)}=\left(\begin{array}{cccc}
E_1 & D_1 & D_2 & D_1 \\ D_1 & E_2 & D_1 & D_2 \\ D_2 & D_1 & E_3 & D_1 \\
D_1 & D_2 & D_1 & E_4
\end{array}\right)
\label{rtH4matrix}  
\end{equation}
where $E_1, E_2, E_3, E_4$ are random energy asymmetries taken from an 
appropriate probability distribution, together with the tunneling parameters 
$D_1$ (n.n. well hopping) and $D_2$ (n.n.n. hopping, $|D_2|\ll |D_1|$). These 
are simple, possible choices; clearly, other special-purpose generalizations 
of the 2LS matrix Hamiltonian are possible and we believe that the 3LS of Eq. 
(\ref{rtH3matrix}) is the minimal generic multiwelled potential wich can take 
the magnetic field into account \cite{Ref14} (the 2LS Hamiltonian of Eq. 
(\ref{rtH2matrix}) could also be adjusted for this purpose, however the energy 
spectrum would be totally insensitive to $B$). One can easily convince oneself, 
at this point, that so long as the energy parameters of the above multiwelled 
effective Hamiltonians obey the usual uniform distribution 
(Eq. (\ref{rtdistrib}), suitably reformulated) as is advocated by the STM, the 
DOS $g(E)$ will remain (roughly) a constant. It is then to be expected that all
these multiwelled local potentials will give rise to the very same physics as 
in the $n_w=2$ case and that thus, in practice, the 2LS choice represents the 
appropriate minimal model for all of the extra low energy excitations 
characterising amorphous solids at low-$T$. It is clear from the above 
discussion, however, that the 2LS tunneling ``particle'' is no atomic particle 
at all, but, on general grounds, it rather represents the local rearrangements 
of a good number of real particles (ions or molecules).
 
\begin{figure}[h]
\includegraphics[width=0.4\textwidth]{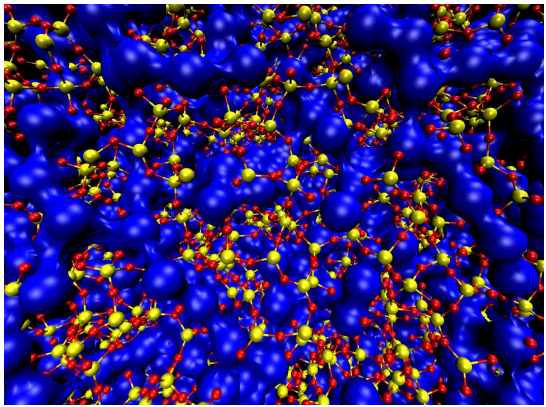} \vskip-1mm
\caption[2]{ (color online) Molecular dynamics snapshot of the structure of 
sodium trisilicate at 2100 K at the density $\rho=2.2$ g cm$^{-3}$. The big 
blue spheres that are connected to each other represent the Na atoms. The Si-O
network is drawn by yellow (Si) and red (O) spheres that are connected to
each other by covalent bonds shown as sticks between Si and O spheres
(from \cite{Ref5}, by permission).
 }
\label{kob}
\end{figure}

\begin{figure}[h]
\includegraphics[width=0.30\textwidth]{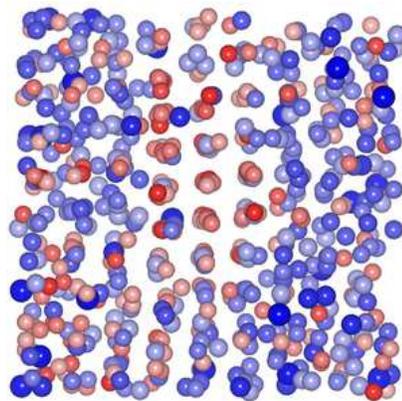} \vskip-5mm
\caption[2]{ (color online) A region including the crystal-like supercluster 
from a snapshot 
of the model simulation -- incorporating fluctuation electron microscopy data 
-- of the Zr$_{50}$Cu$_{45}$Al$_5$ metallic glass at 300 C (from \cite{Ref2}).
The atomic separation distances of the middle zone are about 0.25 nm. This is 
a first realistic image of a crystal embryo in a glass; this object should
not be confused with the concept of short-range order in ideal glasses. }
\label{cembryo}
\end{figure}

All changes if the glass is made up of a mixture of network-forming (NF) ions 
(like those of the good glass-forming SiO$_4$ or (AlO$_4$)$^{-}$ tetrahedral 
groups) {\em as well as} of network-modifying (NM) ions (like those of the good 
crystal-forming K$^{+}$ or Na$^{+}$, or Ba$^{2+}$, ... from the relative 
oxides) which, these last ones, could act as nucleating centres for a partial 
devitrification of the glass, as is known to occur in the multicomponent 
materials \cite{Ref29,Ref30}. Indeed the NM-ions of the good crystal-formers 
are termed  ``glass modifiers'' in the glass chemistry literature \cite{Ref31}
since they do not become part of the interconnected random network but carve
out their own pockets and channels within the glassy network \cite{Ref5,Ref32}.
Fig. \ref{kob} (courtesy from W. Kob, \cite{Ref5}) shows a snapshot of a MD 
simulation of the glass having composition Na$_2$O$\cdot$3(SiO$_2$) (or, 
(Na$_2$O)$_{0.25}$(SiO$_2$)$_{0.75}$) at 2100 K (above $T_g$, in fact) in 
which the non-networking NM Na-atoms are put in evidence (big blue spheres). 
Simulations and experiments in the multisilicates definitely show that the
NM-species in part destroy the networking capacity of the NF-ions and form 
their own clusters inside the NF-network \cite{Ref5}. The chance for these
NM-clusters to be the nest of RER, incipient- or actual micro-crystals is 
obviously very good, considering that these clusters are made of good 
crystal-forming atoms. However, on general grounds and as discussed in the
Introduction, we shall take the attitude that even the purest single-component
glasses will contain RER in some measure. Fig. \ref{cembryo} (from 
\cite{Ref2}) shows one such RER within a snapshot from a joint 
MC-simulation/FEM-measurement on the metallic glass Zr$_{50}$Cu$_{45}$Al$_5$. 
The picture clearly shows an embryo crystal which could not grow to 
macroscopic size due to the arrested dynamics below $T_g$; such structures are 
expected to ubiquitous in all glasses, metallic and non-metallic \cite{Ref33}. 
Except that they are difficult to observe with the available spectroscopic 
tools when sub-nanometric in size. The concentration and size of these RER 
will dictate whether magnetic- or composition-effects become measurable in 
the low-$T$ experiments. $a$-SiO$_2$ in its purest form (Suprasil W) revealed 
no measurable magnetic effects \cite{Ref6,Ref7,Ref8}.

\begin{figure}[ht]
\includegraphics[angle=0, width=0.48\textwidth]{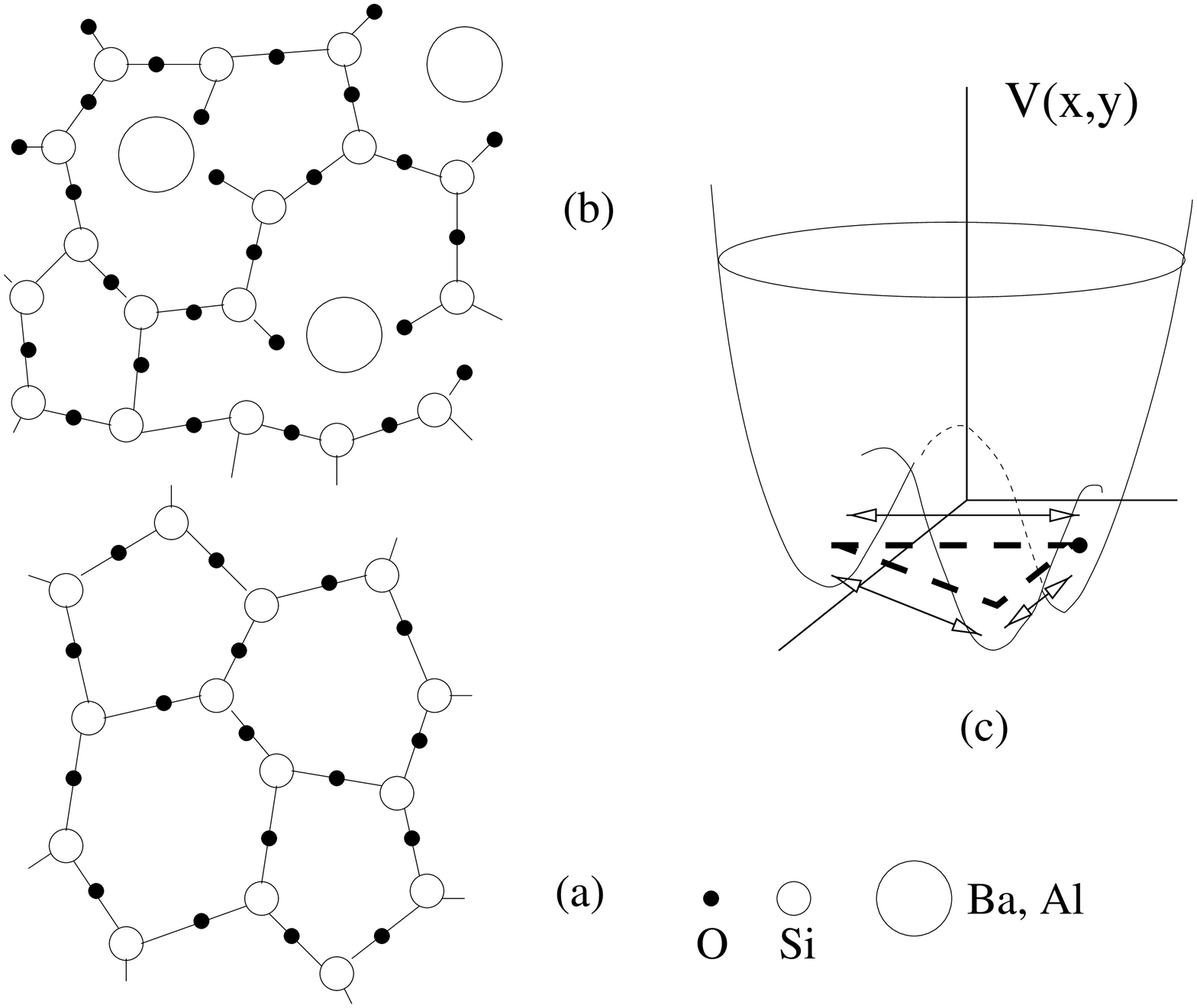}\vskip-5mm
\caption[2]{ Two-dimensional representation of the plausible source of
magnetic-field sensitive (anomalous) tunneling systems in (e.g.) the
AlBaSiO glass. The tight vitreous-SiO$_2$ structure (a) is broken up by 
the Al- and large Ba-atoms (b), thus leaving many metal ions free to move 
in a $n_w$-minima (soft) tunneling potential, with $n\ge3$ (c). The unbroken
Si-O-Si bond dynamics, if any, is of the usual 2LS-type. }
\label{tricone}
\end{figure}

It goes without saying that TS forming in the proximity and within these RER 
or micro-crystalline regions will require a completely different mathematical 
description, in particular the possibility of having more than two wells 
affords a more realistic description of the energy landscape. Hence, $n_w>2$ 
multiwelled systems inside the glass-modifying NM-pockets and -channels should 
follow some new energy-parameters' distribution form when some degree of 
devitrification occurs, leading to entirely new physics. One of the present 
Authors has proposed that precisely this situation occurs inside the 
magnetic-sensitive multicomponent glasses \cite{Ref14}, and in this paper we 
show how this theory explains the $B=0$ composition-dependent dielectric 
and heat capacity data of \cite{Ref18} as well. Instead of the standard 1D 
double-welled (W-shaped) potential, leading to Eq. (\ref{rtH2matrix}), which 
continues to describe the ordinary tunneling 2LS inherent to the 
homogeneously-disordered a-SiO$_2$ network, we take for the TS nested in or 
near the RER, crystal embryos or micro-crystals, the model of a ``particle'' 
having charge $q$ and moving in a $n_w$-welled 3D potential of the shape 
displayed, for $n_w=3$, in Fig. \ref{tricone} for the 2D $(x,y)$-space. The 
hopping 
Hamiltonian of a single, non-interacting tunneling 3LS has therefore the form 
(for a fictitious second-quantization particle in the well-coordinate 
representation) 

\begin{equation}
H_0^{(3)}={\sum}_{i=1}^{3} E_i{c_i}^\dagger c_i + 
{\sum}_{i\neq j}D_0{c_i}^\dagger c_j+{\rm h.c.}
\label{hopping3ls} 
\end{equation}
and is described in matrix form by Eq. (\ref{rtH3matrix}) (where in fact 
$\langle i|H_0^{(3)}|j\rangle$ is displayed, $|i\rangle$ ($i=$1,2,3) denoting 
the single-well ground states). This is our minimal generic model for a 
multiwelled TS. The parameter $D_0$ is chosen positive (contrary to custom 
in the STM, indeed $-\frac{1}{2}\Delta_0< 0$ in Eq. (\ref{rtH2matrix})) 
for a good number of reasons. First, due to the possible softness of the 
local NM-potential, since indeed in general \cite{Ref11} 
$D_0\simeq a\hbar\Omega e^{-bV_0/\hbar\Omega}$, $a$ and $b$ being numbers 
such that for $V_0\gtrsim\hbar\Omega$ $a>0$ and $b=O(1)$ can arise 
\cite{Ref11,Ref14}. This choice is still compatible with the concept of 
tunneling and at the same time yields rather large values of 
$D_0\approx\hbar\Omega$. On more general grounds, however, one should take 
into account that the tunneling ``particle'' is not moving in a vacuum, but 
is embedded in a solid that is for the most part deprived of microscopic 
dynamics, at low-$T$. Thus the surrounding frozen atoms are taking a part in 
the determination of the tunneling particle's lowest stationary states. In the 
case of a perfectly C3-symmetric local $n_w=3$ welled potential of the type 
depicted in Fig. \ref{tricone}, Hamiltonian (\ref{rtH3matrix}) leads to a 
doubly-degenerate ground state and a first excited non-degenerate state (as 
is easily verified from Eq. (\ref{rtH3matrix}) if $E_1=E_2=E_3$). This may 
seem unphysical and yet Sussmann has demonstrated, in a remarkable paper 
\cite{Ref34}, that for electrons trapped in a crystal (or equivalently in a
glass) the situation above described is realised whenever the trapping 
potential is multiwelled with a triangular ($n_w=3$) or tetrahedral ($n_w=4$) 
well-centers geometry. The binding of the seemingly anti-bonding ground state 
is then guaranteed by the TS interaction with the rest of the solid. This 
reasoning is irrelevant for the STM-2LS parameter $\Delta_0$, since both 
positive and negative signs for this parameter yield the same physics. If 
$n_w>2$ the sign will matter and Sussmann's work shows that the choice 
$D_0>0$ is physically justified for an embedded particle in the glass (or 
vacuum). Finally, it will be shown in the Conclusions that in
fact the tunneling ``particle'' cannot be considered a single atom, ion or 
molecule, but rather it represents a cluster of $N$ correlated tunneling 
atomic-scale particles, with $N\approx 200$. Then it is reasonable to expect
that the ground state of such a cluster might be near-degenerate, so our 
choice $D_0>0$ for the effective single tunneling ``particle'' is sound and 
not in conflict with any general quantum-mechanical principle. This $D_0>0$
is the major assumption for the multiwelled TS theory.

\begin{figure}[h]
\includegraphics[width=0.48\textwidth]{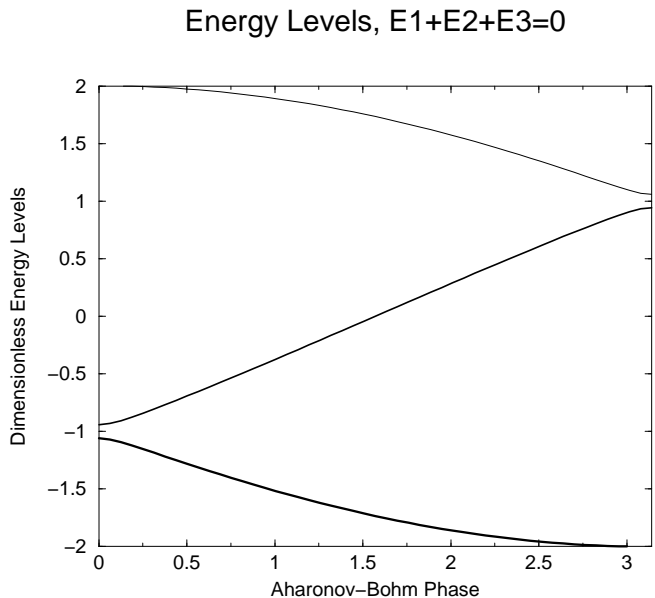} \vskip-5mm
\caption[2]{ Variation with the magnetic Aharonov-Bohm phase $\phi$ of the
energy spectrum (units $D_0=1$) for a choice of $E_1, E_2, E_3$ with
$D/D_0=0.01$. In this work, we are interested in the $\phi=0$ limit of this 
spectrum, which can be treated, at low-$T$, as that of an effective 2LS.}
\label{3lsspectrum}
\end{figure}

At this point, we make a choice for the probability distribution of the 
parameters $E_1, E_2, E_3$ and $D_0$ of a tunneling 3LS nesting in the 
proximity of a RER, crystal embryo or micro-crystal (one could also work with 
a $n_w=4$ model potential, in the Appendix we show that essentially the same 
results can be attained). This is dictated by the fact that near-degeneracy
($E_1=E_2=E_3$) must be favored, yet not fully attained for the wells' energy
asymmetries of one such 3LS. We thus choose, assuming again the tunneling 
potential barriers to be broadly distributed:   

\begin{equation}
P_{ATS}(E_1,E_2,E_3;D_0)=\frac{P^*}{(E_1^2+E_2^2+E_3^2)D_0}.
\label{rtProb3LS}  
\end{equation}
which has the advantage of making use of a dimensionless material-dependent
parameter $P^*$: $P_{ATS}(E_1,E_2,E_3;D_0)$, multiplied by the concentration
$x_{ATS}$ of these anomalous (multiwelled, and now near-degenerate) tunneling 
systems (ATS), is the probability of finding one such ATS per unit volume. In 
the following, $x_{ATS}$ will be absorbed in the parameter $P^*$. This choice 
for $P_{ATS}$ has provided a good description of the experimental data for the 
multisilicates in a magnetic field \cite{Ref14}, when in the Hamiltonian 
(\ref{rtH3matrix}) (or equivalently (\ref{hopping3ls})) $D_0$ at position 
$(i,j)$ is replaced with $D_0e^{i\phi_{ij}}$ ($\phi_{ij}$ being the 
appropriate Peierls phase). As was shown in Ref. \cite{Ref14}, the spectrum 
of this $B>0$ modified 3LS Hamiltonian (\ref{rtH3matrix}) is formally given 
by (using Vi\`ete's formula for the cubic equation's solutions):

\begin{eqnarray}
&&\frac{{\cal E}_k}{D_0}
= 2\sqrt{ 1-\frac{\sum_{i\not=j}E_iE_j}{6D_0^2} } ~
\cos(\frac{1}{3}\theta+\theta_k)
\label{spec3ls} \\
&&\cos\theta = \left( \cos\phi+\frac{E_1E_2E_3}{2D_0^3} \right)
\left( 1-\frac{\sum_{i\not=j}E_iE_j}{3D_0^2} \right)^{-3/2},
\nonumber
\label{spectrum}
\end{eqnarray}
\noindent
(with $k$=0, 1, 2 and $\theta_k=0,+\frac{2}{3}\pi,-\frac{2}{3}\pi$
distinguishing the three lowest eigenstates) and for a choice of 
$E_1, E_2, E_3$ and $D_0\gg \sqrt{E_1^2+E_2^2+E_3^2}$ (near-degenerate limit) 
this is shown in Fig. \ref{3lsspectrum}. One can see that for very small 
$\phi$ (Aharonov-Bohm phase, proportional to the magnetic field $B$) the 
spectrum consists of an isolated near-degenerate doublet which is well 
separated from the higher excited states. We shall exploit the $\phi=0$ limit 
of this description for an explanation of the composition-dependent 
experiments.

It should be stressed at this point that in the absence of a magnetic field, 
like in this work, one could make use of a 2LS minimal model for the 
description of the ATS, $H_0^{(2)}(E_1,E_2;D_0)$, and with the distribution 
$P(E_1,E_2;D_0)=P^*/D_0\sqrt{E_1^2+E_2^2}$ ensuing from the proximity of 
RER or incipient micro-crystallites. It was shown in Ref. \cite{Ref14} 
that, at least for the heat capacity, this leads to the same physics as 
obtained from the 3LS multiwelled model. There is no harm in using, for the 
ATS nesting in the incipient crystalline regions, a more realistic minimal 
generic multiwelled model like the above 3LS Hamiltonian $H_0^{(3)}$ which 
better approximates the physical reality of the energy landscape. Moreover, 
the model for the composition-dependent effects remains the very same used for 
the magnetic effects and many results already obtained for that theory can be
exploited by setting simply $B=0$. We remark, also, that a distribution of the 
type (\ref{rtProb3LS}) for the energy asymmetry was already proposed for the 
explanation of low-$T$ experiments with mesoscopic Au and Ag wires 
\cite{Ref36}, where TS (of standard 2LS type) were advocated and where the
poly-crystallinity of metals must be accounted for.  

\begin{figure}[ht]
\includegraphics[angle=0, width=0.60\textwidth]{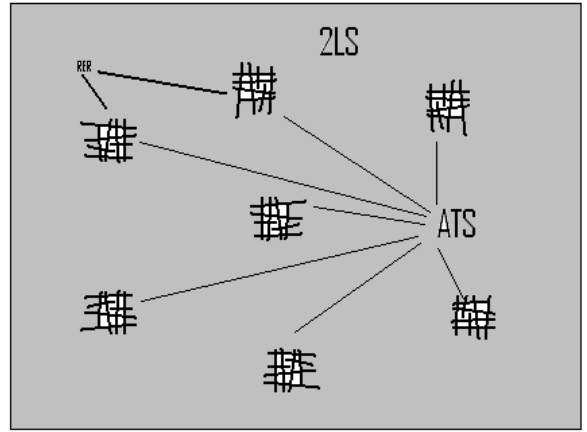}\vskip-5mm
\caption[2]{A 2D cartoon of the chocolate-like, ceramic-glass structure of a 
real glass, in which partial devitrification has occurred, with the location of 
its low-$T$, two-species TS. In the randomly-networked bulk of the material sit
the STM-2LS, with their own concentration $x_{2LS}$, whilst within and in the 
proximity of the incipient crystallites nest the ATS, with their own bulk 
concentration $x_{ATS}$, each being described by Eqs. (\ref{rtH3matrix}) and 
(\ref{rtProb3LS}). We expect $x_{ATS}< x_{2LS}$ and that $x_{ATS}\to 0$
in the best glasses.}    
\label{cartoon}
\end{figure}

In summary, we have fully justified the extended TM which we have used in Ref. 
\cite{Ref14} and which we exploit also in this paper. The realistic glass is 
recognized to have a structure resembling that of chocolate \cite{Ref35} and 
as is pictured in the cartoon in Fig. \ref{cartoon}: a 
homogeneously-disordered networked solid in which (at low-$T$ in the glass) 
only standard 2LS are present with their own concentration $x_{2LS}$ and in 
which incipient crystallites are embedded. (For chocolate, these would be 
sugar crystals). In the proximity or within these crystallites are nested the 
ATS, with their own concentration $x_{ATS}$ in the solid and with their own 
quantum mechanics and statistics defined by the minimal generic model 
represented by Eqs. (\ref{rtH3matrix}) and (\ref{rtProb3LS}). This is by no 
means an ad-hoc model, since the very same model would describe TS in  all 
types of real metallic and non-metallic glasses and quantitatively explains 
all of the low-$T$ experiments in non-metallic glasses tackled so far.

\section{III. PREDICTIONS FOR THE DIELECTRIC CONSTANT}

The 2LS-STM has been successful in the semi-quantitative explanation of a 
variety of interesting thermal, dielectric and acoustic anomalies of 
structural glasses at temperatures $T<$ 1 K \cite{Ref10,Ref11}, the physics 
of cold glasses being important not only for its universalities, but also 
because of its link with the physics of the glass transition (see, 
e.g., \cite{Ref37}). Beside the linear in $T$ behavior of the heat capacity 
$C_p$, it is believed that the linear in $\pm\ln T$ behavior of the real-part 
of the frequency-dependent dielectric constant $\epsilon'(T,\omega)$ 
represents a cogent characterization of the glassy state at low temperatures. 
We begin by deriving this behavior and putting it to the test on data for 
$\epsilon'$ for pure amorphous silica (no measurable ATS effects, that is).

In the presence of an applied electric field ${\bf F}$ we must add the dipole 
energy $-{\bf F}\cdot{\bf p}_0$ to the parameter $\frac{1}{2}\Delta$ in the 
expression (\ref{rtH2matrix}) for the low-energy Hamiltonian $H_0^{(2)}$. We 
can express the permittivity as
$\epsilon=-\left. \frac{\partial^2 f(F)}{\partial F^2}\right|_{F=0}$, where 
$f(F)=-\frac{1}{k_BT}\ln Z(F)$ represents the free energy per unit volume. 
The statistical average implies also an integration over the two parameters 
of the 2LS, $\Delta$ and $\Delta_0$, according to the distribution given by 
Eq. (\ref{rtdistrib}). We can write the partition function in terms of the
energy levels ${\cal E}_{1,2}$:  
$Z=e^{-{\cal E}_{1}/k_BT}+e^{-{\cal E}_{2}/k_BT}$.

\begin{figure}[h]
\includegraphics[width=0.48\textwidth]{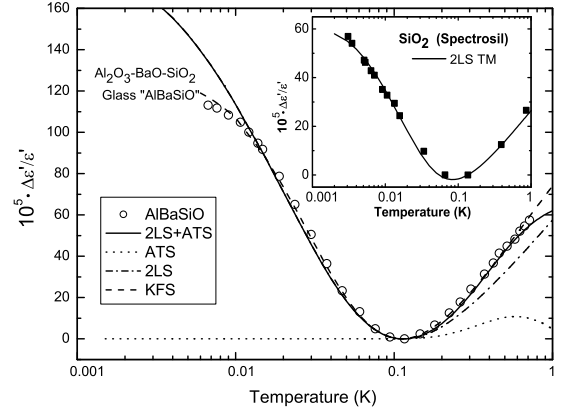} \vskip-5mm
\caption[2]{ Dielectric signature of pure $a$-SiO$_2$ (inset) and
AlBaSiO (main) glasses. SiO$_2$ data \cite{Ref39}, fitted with Eq.
(\ref{dc2ls}), display a -2:1 2LS TM behavior. AlBaSiO data
\cite{Ref40} display rather a -1:1 behavior, yet could be fitted
with Eq. (\ref{dc2ls}) (dashed line) \cite{Ref40} with a large
$\Delta_{0min}=$12.2 mK 2LS tunneling parameter. We have fitted
all data with a more realistic $\Delta_{0min}=$3.9 mK and best fit
parameters from Table \ref{tabl1} using Eqs. (\ref{dc2ls}) and 
(\ref{dcats}) (driving frequency $\omega=$1 kHz). } 
\label{fig1}
\end{figure}

Fig. \ref{fig1} (inset) shows the behavior of the $T$-dependent part of
$\epsilon'(T,\omega)$,
$\Delta\epsilon'/\epsilon'=[\epsilon'(T)-\epsilon'(T_0)]/\epsilon'(T_0)$,
(where $T_0(\omega)$ is a characteristic minimum) for pure vitreous SiO$_2$
(Spectrosil). It can be seen that linear regimes in  $-\ln T$ for $T<T_0$ 
and $+\ln T$ for $T>T_0$ are observed, and roughly with slopes $S_{-}=-2S$ 
and $S_{+}=+S>0$, or in a -2:1 ratio. According to the 2LS-STM, in fact, we 
have the expressions \cite{Ref10,Ref11,Ref38}

\begin{eqnarray}
&&\left. \frac{\Delta\epsilon'}{\epsilon'} \right|_{2LS}= \left.
\frac{\Delta\epsilon'}{\epsilon'} \right|_{2RES}
+\left. \frac{\Delta\epsilon'}{\epsilon'} \right|_{2REL}, 
\label{dc2ls} \\
&&\left. \frac{\Delta\epsilon'}{\epsilon'} \right|_{2RES}=
\frac{2\bar{P}\overline{p_0^2}}{3\epsilon_0\epsilon_r}
\int_{z_{min}}^{z_{max}}\frac{dz}{z} \sqrt{1-\left(
\frac{\Delta_{0min}}{2k_BTz} \right)^2}\tanh z,
\nonumber \\
&&\left. \frac{\Delta\epsilon'}{\epsilon'} \right|_{2REL}=
\frac{\bar{P}\overline{p_0^2}}{3\epsilon_0\epsilon_r}\times \nonumber \\
&&\times\int_{z_{min}}^{z_{max}}dz~\int_{\tau_{min}}^{\tau_{max}}
\frac{d\tau}{\tau}\sqrt{1-\frac{\tau_{min}}{\tau}}
\cosh^{-2}(z)\frac{1}{1+\omega^2\tau^2}, \nonumber
\end{eqnarray}
where we neglect (for low $\omega$) the frequency-dependence in the RES part, 
where $z_{min,max}=\Delta_{0min,max}/2k_BT$ and where $\tau$ is the 
phenomenological 2LS relaxation time given by (with $E=2k_BTz$) \cite{Ref11}
\begin{equation}
\tau^{-1}=E\Delta_0^2/\gamma\tanh\left( \frac{E}{2k_BT}\right ).
\label{rt2ls}
\end{equation}

\noindent
In these expressions, $\Delta_{0min}$ and $\Delta_{0max}$ are $\Delta_0$'s  
phenomenological bounds, $\gamma$ is an elastic material parameter of the 
solid and $\tau_{min}^{-1}=E^3/\gamma\tanh\left( \frac{E}{2k_BT} \right)$,
$\tau_{max}^{-1}=E\Delta_{0min}^2/\gamma\tanh\left(\frac{E}{2k_BT} \right)$. 
$\bar{P}$ (containing the 2LS volume concentration, $x_{2LS}$) is the 
probability per unit volume and energy that a 2LS occurs in the solid (it 
appears in Eq. (\ref{rtdistrib})) and $\overline{p_0^2}$ is the average 
square 2LS electric dipole moment. Moreover, the strategy of dielectric 
relaxation theory has been adopted, whereby the full complex dielectric 
constant $\epsilon(T,\omega)$ has been written as, for $\omega\tau\ll 1$ 
\cite{Ref38,Ref41}

\begin{equation}
\epsilon(T,\omega)=\epsilon'_{RES}(T)+\epsilon'_{REL}(T)
\frac{1}{1+i\omega\tau}
\label{relax}
\end{equation}
\noindent
the subfixes $RES$ and $REL$ referring to the zero relaxation-time resonant 
and, respectively, relaxational contributions to the linear response 
$\epsilon'$ at zero frequency. 

Presently, from expressions (\ref{dc2ls}) we deduce that: 1) The so-called 
resonant (RES) contribution has the leading behavior
\begin{eqnarray}
\left. \frac{\Delta\epsilon'}{\epsilon'}
\right|_{2RES}\simeq\cases{
-\frac{2}{3}\frac{\bar{P}\overline{p_0^2}}{\epsilon_0\epsilon_r}
\ln\left( \frac{2k_BT}{\Delta_{0max}} \right) & if
$T<\frac{\Delta_{0max}}{2k_B}$,\cr 0 & if
$T>\frac{\Delta_{0max}}{2k_B}$;\cr }
\end{eqnarray}
2) the relaxational (REL) contribution has, instead, the leading behavior
\begin{eqnarray}
\left. \frac{\Delta\epsilon'}{\epsilon'}
\right|_{2REL}\simeq\cases{ 0 & if $\omega\tau_{min}\gg1$ \cr
\frac{1}{3}\frac{\bar{P}\overline{p_0^2}}{\epsilon_0\epsilon_r}
\ln\left( \frac{2k_BT}{\Delta_{0min}} \right) & if
$\omega\tau_{min}\ll 1$. \cr }
\end{eqnarray}
Thus, the sum of the two contributions has a V-shaped form, in a 
semilogarithmic plot, with the minimum occurring at a $T_0$ roughly given by 
the condition $\omega\tau_{min}(k_BT)\simeq 1$, or 
$k_BT_0(\omega)\simeq (\frac{1}{2}\gamma\omega)^{1/3}$.
$\epsilon_0\epsilon_r$ is here the bulk of the solid's dielectric constant 
and we see that a -2:1 characteristic behavior is justified by the STM with 
the $T>T_0$ slope given by $S=\bar{P}\overline{p_0^2}/3\epsilon_0\epsilon_r$.

\begin{figure}[h]
\includegraphics[width=0.48\textwidth]{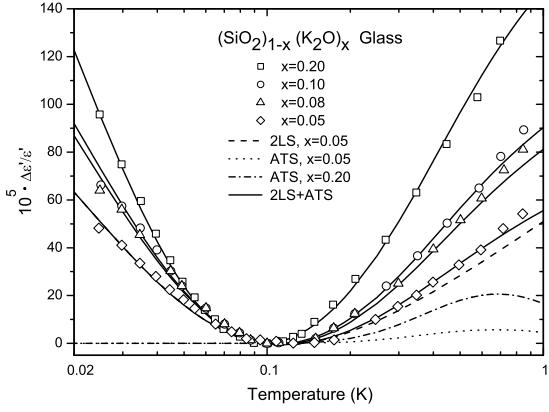} \vskip-5mm
\caption[2]{ Dielectric signature of mixed
(SiO$_2$)$_{1-x}$(K$_2$O)$_x$ glasses as function of $T$ and $x$
\cite{Ref18}. Fitting parameters from Table \ref{tabl1} using Eq.
(\ref{dc2ls}) and (\ref{dcats}) from our theory (driving frequency
$\omega=$10 kHz). }
 \label{fig2}
\end{figure}

This behavior is observed in pure $a$-SiO$_2$ \cite{Ref39} (Fig. \ref{fig1} 
(inset), with the fitting parameters of Table \ref{tabl1}, $x=0$, from our 
own best fit to Eq. (\ref{dc2ls})). However, in most multicomponent glasses 
one more often observes a V-shaped curve with a (roughly) -1:1 slope ratio. 
Fig. \ref{fig1} (main) shows this phenomenon for the multisilicate AlBaSiO 
glass (in fact, a MAS-type ceramic-glass), which has been extensively 
investigated in recent times due to its unexpected magnetic field response 
\cite{Ref6,Ref7,Ref8,Ref14}. Also, Fig. \ref{fig2} shows the remarkable 
behavior of the dielectric constant vs. $T$ for the glasses of composition 
(SiO$_2$)$_{1-x}$(K$_2$O)$_x$ containing a molar concentration $x$ of alkali 
oxide \cite{Ref18}. It is seen that a $S_{-}/S_{+}$ slope ratio of roughly 
-1:1 is observed, with the slope definitely changing with $x$ (and faster for 
$T>T_0$). These data from the Anderson group \cite{Ref18}, thus far 
unexplained by the 2LS-STM, call for an extension of the accepted STM and we 
show below that a simple explanation can be given in terms of the very same 
ATS that have been justified in Section II and advocated by one of us in order 
to explain the magnetic response of AlBaSiO and other multicomponent glasses 
\cite{Ref14}. In view of the interest for these materials in low-$T$ 
metrology, and on fundamental grounds, such explanation appears overdue to 
us. Moreover,``additional'' TS (beside the standard 2LS) of the type here 
advocated were already called for in \cite{Ref18} and other theoretical 
papers \cite{Ref39b}.

For the multiwelled (3LS, in practice) Hamiltonian (\ref{rtH3matrix}) we have 
$n_w=3$ low-lying energy levels, with ${\cal E}_0<{\cal E}_1\ll{\cal E}_2$. 
In the $E_i\rightarrow0$ and $D\equiv\sqrt{E_1^2+E_2^2+E_3^2} \ll D_0$ limits
(due to the chosen near-degenerate distribution, Eq. (\ref{rtProb3LS})) we 
can approximate the $n_w=3$ - eigenstate system through an {\em effective 2LS} 
(though sensitive to all three well-asymmetries and their distribution) having 
gap $\Delta{\cal E}={\cal E}_1-{\cal E}_0$:

\begin{equation}
\lim\Delta{\cal E}\simeq\sqrt{E_1^2+E_2^2+E_3^2}\equiv D.
\label{rtDeltaEPS}  
\end{equation}
\noindent
(we have also exploited the condition $E_1+E_2+E_3=0$). Using the theory of 
\cite{Ref14} to work out the 3LS-contributions to $\epsilon'_{RES}$ and 
$\epsilon'_{REL}$, we arrive at the following expressions for the 
contribution to the dielectric anomaly from the advocated ATS:
\begin{eqnarray}
&&\left. \frac{\Delta\epsilon'}{\epsilon'} \right|_{ATS}= \left.
\frac{\Delta\epsilon'}{\epsilon'} \right|_{ARES}
+\left. \frac{\Delta\epsilon'}{\epsilon'} \right|_{AREL}, \label{dcats} \\
&&\left. \frac{\Delta\epsilon'}{\epsilon'} \right|_{ARES}=
\frac{\pi\tilde{P}^*\overline{p_1^2}}{3\epsilon_0\epsilon_rD_{min}}
\int_1^{\infty}\frac{dy}{y^2}\tanh\left( \frac{D_{min}}{2k_BT} y
\right),
\nonumber \\
&&\left. \frac{\Delta\epsilon'}{\epsilon'} \right|_{AREL}=
\frac{\pi\tilde{P}^*\overline{p_1^2}}{2\epsilon_0\epsilon_rD_{min}}
\left( \frac{D_{min}}{2k_BT} \right)\times
\nonumber \\
&&\times\int_1^{\infty}\frac{dy}{y}\cosh^{-2}\left(
\frac{D_{min}}{2k_BT} y \right)\frac{1}{1+\omega^2\tau_{Amax}^2}.
\nonumber
\end{eqnarray}
Here we have again neglected, for low-$\omega$, the frequency-dependence in 
the RES part, we have put $y=D/D_{min}$ and $\tau_{Amax}$ is the largest 
phenomenological ATS relaxation time given by \cite{Ref42}
\begin{equation}
\tau_{Amax}^{-1}=D^5/\Gamma\tanh\left( \frac{D}{2k_BT}\right ).
\label{rtats}
\end{equation}
Moreover $D_{min}$ is the lowest energy gap of the multilevel ATS, $\Gamma$ 
is another appropriate elastic constant and $\tilde{P}^*$ is the (slightly 
renormalised) probability per unit volume (after inclusion of $x_{ATS}$) 
that an ATS occurs within the NM-pockets and channels, with $\overline{p_1^2}$
the average square ATS dipole moment. $\tilde{P}^*$ and $P^*$ are so related:
 
\begin{equation}
\tilde{P}^*=P^*\ln\left(\frac{D_{0max}}{D_{0min}}\right)
\end{equation}
$D_{0min}$ and $D_{0max}$ being $D_0$'s lower and upper bounds, respectively.
This description is intimately linked to the chosen distribution function, 
Eq. (\ref{rtProb3LS}), for these ATS which is favoring near-degenerate energy 
gaps $D$ bound from below by $D_{min}$. In turn, this produces an overall 
density of states given by (\cite{Ref14}, for $B=0$): 

\begin{equation}
g(E)=g_{2LS}+g_{ATS}(E)\simeq 2\bar{P}+\frac{2\pi\tilde{P}^*}{E}
\theta(E-D_{min})
\label{dos}
\end{equation}
and that is now roughly of the form advocated by Yu and Legget \cite{Ref19} 
and by some other preceeding Authors (e.g. \cite{Ref43}) to explain 
anomalies not accounted for by the standard 2LS TM. $\theta(x)$ is the step
function.

Manipulation of the expressions in (\ref{dcats}) shows that: 1) The RES 
contribution from the ATS has the leading behavior (note that for
$T<D_{min}/2k_B$, $\epsilon'|_{ARES}$ is roughly a constant)
\begin{eqnarray}
\left. \frac{\Delta\epsilon'}{\epsilon'}
\right|_{ARES}\simeq\cases{ 0 & if $T<\frac{D_{min}}{2k_B}$, \cr
\frac{\pi\tilde{P}^*\overline{p_1^2}}{6\epsilon_0\epsilon_rk_BT}
\ln\left( \frac{2k_BT}{D_{min}} \right) & if
$T>\frac{D_{min}}{2k_B}$; \cr }
\end{eqnarray}
2) the REL contribution is, instead, characterised by the leading
form
\begin{eqnarray}
\left. \frac{\Delta\epsilon'}{\epsilon'}
\right|_{AREL}\simeq\cases{ 0 & if $\omega\tau_{Amax}\gg1$ \cr
\frac{\pi\tilde{P}^*\overline{p_1^2}}{\epsilon_0\epsilon_rk_BT}
\ln\left( \frac{k_BT}{D_{min}} \right) & if $\omega\tau_{Amax}\ll
1$. \cr }
\end{eqnarray}
Thus, the V-shaped semilogarithmic curve is somewhat lost. However, adding 
the 2LS (Eq. (\ref{dc2ls})) and ATS (Eq. (\ref{dcats})) contributions 
together one does recover a rounded semilog V-shape with a slope
$S_{-}\simeq -2S$ basically unchanged for $T<T_0$ and an augmented
slope $S_{+}=S+S_{ATS}$ for $T>T_0$ with
$S_{ATS}=7\pi\tilde{P}^*\overline{p_1^2}/6\epsilon_0\epsilon_rk_BT$
that for $T<D_{min}/k_B$ may approach $2S$ and thus (qualitatively) explain
a -1:1 slope ratio.

\begin{figure}[h]
\includegraphics[width=0.48\textwidth]{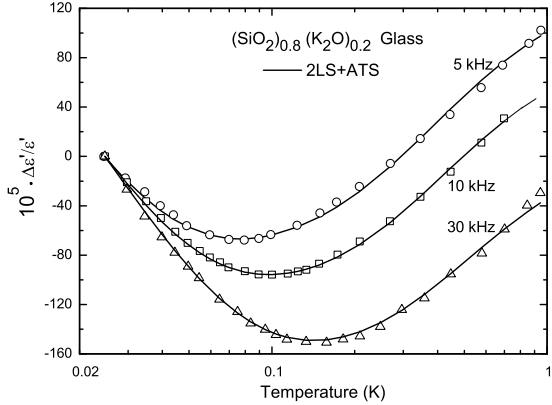} \vskip-5mm
\caption[2]{ Dielectric signature of mixed
(SiO$_2$)$_{1-x}$(K$_2$O)$_x$ glasses as function of $T$ and $\omega$ for 
$x=$0.2 \cite{Ref18}. Fitting parameters from Table \ref{tabl1} using Eq. 
(\ref{dc2ls}) and (\ref{dcats}) from our theory. }
\label{fig3}
\end{figure}

\begin{table}[h]
\begin{tabular}{|l|cccccc|}
\hline glass & $x$ & $A_{2LS}$ & $\gamma$ & $A_{ATS}$ &
$D_{min}$ & $\Gamma$ \\
type & mol & 10$^{-5}$ & 10$^{-8}$ sJ$^3$ & 10$^{-5}$ & K &
10$^{-6}$ sK$^5$ \\
\hline \hline
SiO$_2$ & 0 & 47.2  & 5.30 & -  & -  & - \\
\hline
AlBaSiO & - & 116.2 & 13.40 & 264.7 & 0.65 & 69.73 \\
\hline
K-Si & 0.05 & 104.1 & 1.33 & 75.5 & 0.87 & 3.55 \\
\hline
K-Si & 0.08 & 146.5 & 1.23 & 130.0 & 0.87 & 3.97 \\
\hline
K-Si & 0.10 & 158.5 & 1.15 & 160.0 & 0.87 & 5.08 \\
\hline
K-Si & 0.20 & 239.5 & 0.82 & 281.9 & 0.87 & 6.44 \\
\hline
\end{tabular}
\caption[1]{ Extracted parameters for the glasses; K-Si stands for
the (SiO$_2$)$_{1-x}$(K$_2$O)$_x$ glasses. In all of the best fits
we have employed the values $\Delta_{0min}=$3.9 mK and
$\Delta_{0max}=$10 K extracted from fitting the pure SiO$_2$ data
of Fig. \ref{fig1} (inset).} 
\label{tabl1}
\end{table}

\begin{figure}[h]
\includegraphics[width=0.48\textwidth]{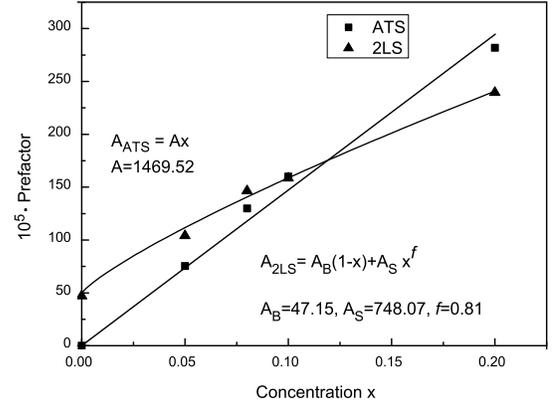} \vskip-5mm
\caption[2]{ The 2LS and ATS dimensionless prefactor parameters 
($\times 10^5$) for all glasses (from Table \ref{tabl1}) as a function of 
$x$. Our data fit well with our theoretical expectations (full lines). }
\label{fig4}
\end{figure}

We have fitted the full expressions (\ref{dc2ls}) and (\ref{dcats}) to the 
data for AlBaSiO in Fig. \ref{fig1} (main) and to the $x$-dependent data for 
(SiO$_2$)$_{1-x}$(K$_2$O)$_x$ in Figs. \ref{fig2} and \ref{fig3}, obtaining 
in all cases very good agreement between theory and experiments \cite{Ref18}. 
Fig. \ref{fig3} shows the fit of our theory to the frequency-dependent data 
for $x=0.2$. In all of these best fits we have kept the value of 
$\Delta_{0min}=3.9$ mK fixed, as obtained from our pure SiO$_2$ fit, and the 
value of $D_{min}$ also independent of $x$ and $\omega$. The idea is that 
these parameters are rather local ones and should not be influenced by NF/NM
dilution. Table \ref{tabl1} gathers the values of all the (2LS and ATS) 
parameters used for our best fits and Fig. \ref{fig4} shows the dependence 
of the prefactors (containing $x_{2LS}$ in $\bar{P}$ and $x_{ATS}$ in 
$\tilde{P}^*$) with $x$. It can be seen that, as expected, the ATS prefactor
$A_{ATS}=\pi\tilde{P}^*\overline{p_1^2}/\epsilon_0\epsilon_rD_{min}$
scales linearly with $x$, an excellent confirmation that the ``additional'' 
TS of \cite{Ref18,Ref39b} are those ATS, by us proposed and modelled as 3LS, 
forming near and inside the microcrystallites that may nucleate within the 
NM-pockets and channels. It can be seen, instead, that the 2LS prefactor
$A_{2LS}=\bar{P}\overline{p_0^2}/\epsilon_0\epsilon_r$ of our fits also 
increases, though less rapidly, with increasing $x$ (a decrease like $1-x$ 
would be expected). We propose (adopting a NF-, NM-cluster percolation 
picture) that new, ``dilution-induced'' 2LS form with alkali mixing near the 
NF/NM interfaces of the NF percolating cluster(s) as $x$ is increased from 0. 
This reasoning leads to the expression $A_{2LS}=A_{bulk}(1-x)+A_{surf}x^f$ 
for the 2LS prefactor, with $A_{bulk}$, $A_{surf}$ and $f$ fitting 
parameters. Our best fit leads to the value $f=0.81$, in fair agreement with 
the euristic expression

\begin{figure}[h]
\includegraphics[width=0.50\textwidth]{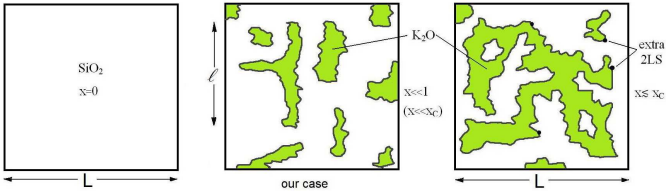} \vskip-5mm
\caption[2]{ (color online) A cartoon of the fractal (presumably percolating) 
geometry of the NM-pockets and channels (green), these NM-clusters growing 
with increasing $x$.}
\label{percolation}
\end{figure}

\begin{equation}
f=1-(D-D_s)\nu
\label{fractal} 
\end{equation}
(where $D$ is the fractal dimension of the percolating cluster, $D_s$ with 
$D_s\le D$ is that of its ``bridging'' surface (not necessarily the hull) 
and $\nu$ is the connectedness length's exponent) that one would deduce from 
elementary fractal or percolation theory (see, e.g. \cite{Ref43b}). $D_s$
is the fractal dimension of that part of the NM random-cluster's surface where 
formation of TS takes place and we expect $2\le D_s\le D$. It is indeed 
reasonable to expect new TS to be forming at these NM/NF random interfaces, 
for these are surfaces of chemical discontinuity in the material. The above 
expression is derived as follows. 
Imagine (as is shown in the cartoons in Fig. \ref{percolation}) the 
NM-clusters percolating through the NF-bulk with a site concentration $x$, so 
that their volume scales like
${\cal V}\sim \ell^D$ where $\ell\sim x^{\nu}$ is their typical linear 
size. The number of 2LS on the surface of these clusters will scale like 
$N_{2LS}^{(s)}\sim x \ell^{D_s}$ and so their density like
$N_{2LS}^{(s)}/{\cal V}\sim x x^{(D_s-D_f)\nu}=x^f$ with the given
expression, Eq. (\ref{fractal}), for $f$. If we consider clusters of 2D 
percolation and assume $D_s=D_h=7/4$ (the fractal dimension of the hull of
the spanning cluster), then with $D=91/48$ and $\nu=4/3$ \cite{Ref43b} we 
would get $f=29/36=0.8055$. More realistically, on the assumption of 
percolating 3D NM-clusters in the mixed glasses, we can make use of the 
values \cite{Ref43b, Ref43c} $D\simeq 2.52$, $D_s=D_h=2.14$ and 
$\nu\simeq 0.88$ to arrive at the value $f=0.67$ using Eq. (\ref{fractal}) 
\cite{Ref43d}. It is however not at all clear where, at the NM/NF fractal 
interfaces, the new 2LS will form (i.e. what the exact definition of $D_s$ 
ought to be: hull surface sites, screening sites, dead-end sites ...). If all
of the hull sites are involved, then for 3D $x=x_c$ percolation $D_s=D$ and 
one then expects $f=1$. Thus, this new phenomenology opens a tantalizing new 
investigation avenue for research on the applications of fractal theory to 
low-$T$ physics. At the same time, knowledge of which type of NM/NF fractal 
interface sites are involved in the TS-formation would greatly improve our 
understanding about the microscopic nature of the TS (see also Ref. 
\cite{Ref21}).

\section{IV. PREDICTIONS FOR THE HEAT CAPACITY}

\begin{figure}[h]
\includegraphics[width=0.48\textwidth]{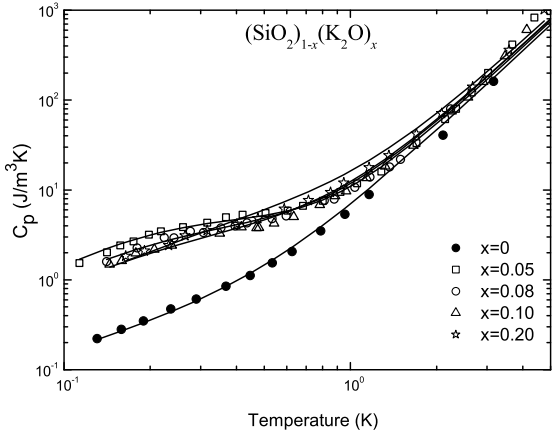} \vskip-5mm
\caption[2]{ The temperature dependence of the heat capacity for a-SiO$_{2}$ 
(black circles) and for the (SiO$_2$)$_{1-x}$(K$_2$O)$_x$ glasses 
\cite{Ref18}. The full lines are our theoretical curves, as generated by 
Eq.(\ref{rtCpSum}). }
\label{fig5}
\end{figure}

We now come to the explanation of the, also rather anomalous, heat-capacity 
data for the mixed glasses (SiO$_2$)$_{1-x}$(K$_2$O), reported in \cite{Ref18}
as a function of $T$ and for different $x$. The heat capacity's 
low-temperature dependence in zero magnetic field is, for pure glasses, 
usually given by the following expression:

\begin{equation}
C_p\left(T\right)=B_{ph}T^3\ +B_{2LS}T.
 \label{rthcapac}
\end{equation}
The first term accounts for the Debye-type contribution from the acoustic 
phonons and dominates above 1 K, the second term is usually attributed to the 
low-energy excitations specific of all vitreous solids - the tunneling 2LS. 
$B_{ph}$ and $B_{2LS}$ are material-dependent constants. This expression 
describes well the experimental data for pure silica glass at zero field 
(Fig. \ref{fig5}, black circles: $x$=0 with fit parameters from Table 
\ref{tabl2}), but it fails for the multicomponent glasses, like AlBaSiO, BK7, 
Duran (see e.g. \cite{Ref14} and references therein) and for the mixed glasses
(SiO$_2$)$_{1-x}$(K$_2$O)$_x$ for $x>0$ \cite{Ref18}.

Typically, the heat capacity's experimental data for the multicomponent 
glasses in zero field denote a kind of ``shoulder'' at intermediate-low 
temperatures. This suggests a density of states, for at least some of the 
independent TS in the glass, of the form $g(E)\propto 1/E$, in contrast to 
the standard 2LS-TM prediction, $g(E)\simeq$ const., which ensues from the 
standard TM distribution of parameters. Indeed, this $1/E$ contribution to 
the DOS was the very first observation that has led to the hypothesis of the 
ATS formulated in \cite{Ref14}.

To find out the precise expression for the heat capacity due to the ATS we 
make use of the 3LS formulation for the ATS described in \cite{Ref14} and in 
more detail in Section II. The heat capacity is determined from the second 
derivative of the free energy with respect to temperature:

\begin{equation}
C^{ATS}_p(T)=-T\frac{{\partial}^2F_{ATS}(T)}{\partial T^2},
\label{rtCAts}
\end{equation}
where $F_{ATS}(T)$ is the free energy of the ATS given by, if we  neglect the 
third, highest energy level in the spectrum of Hamiltonian (\ref{rtH3matrix}) 
(effective 2LS approximation):

\begin{eqnarray}
&& F_{ATS}(T)=-k_BT\ln{\left(e^{-\frac{{\cal E}_0}{k_BT}}
+e^{-\frac{{\cal E}_1}{k_BT}}\right)}
 \nonumber \\
&& =-k_BT\ln{\left(2{\cosh\left(\frac{E}{2k_BT}\right)}\right)},
\label{rtFAts}
\end{eqnarray}
with $E={\cal E}_1-{\cal E}_0$. The heat capacity is then obtained by 
averaging over the parameter distribution, or, equivalently, by a convolution 
with the DOS:

\begin{equation}
C^{ATS}_p(T)=k_B\int^{\infty}_0 dE g_{ATS}(E)
{\left(\frac{E}{2k_BT}\right)}^2\cosh^{-2}\left(\frac{E}{2k_BT}\right),
\label{rtCAtsEv}
\end{equation}
where density of states $g_{ATS}(E)$ has the following form \cite{Ref14}:

\begin{eqnarray}
&& {\rm g}_{ATS}(E)=\int dD\int {dD}_0 P(D,D_0)\delta(E-D)\simeq  
\nonumber \\
&& \simeq\cases{ \frac{2P^*}{E} & if $E>D_{min}$, \cr 0 & if $E<D_{min}$; \cr }
\label{rtgATS}
\end{eqnarray}
and $D_{min}$ is the lower cutoff. The final expression for the ATS heat 
capacity results in \cite{Ref14}:

\begin{eqnarray}
&& C^{ATS}_p(T)=B_{ATS}\left[ 
\ln\left(2\cosh\left(\frac{D_{min}}{2k_BT}\right)\right)-\right.
\nonumber \\
&&  \left.-\frac{D_{min}}{2k_BT}\tanh\left(\frac{D_{min}}{2k_BT}\right) \right]
\label{rtCAtsfinal}
\end{eqnarray}
where the prefactor for the ATS is 
$B_{ATS}=2\pi \tilde{P}^*k_Bn_{ATS}\rho(x)$, $\tilde{P}^*$ as in Section III, 
$n_{ATS}$ being the ATS mass concentration, $\rho(x)$ the glass' mass density.
Of course, $x_{ATS}=n_{ATS}\rho(x)$.
For $k_BT\gtrsim D_{min}$ this is indeed roughly a constant and gives the 
observed ``shoulder'' in $C_{p}(T)$ when the contribution $B_{ph}T^3$ (from 
virtual phonons) as well as the STM linear term $B_{2LS}T$ are taken into
account.

Both prefactors, for the 2LS and ATS contributions, are dependent on the molar 
concentration $x$ of alkali-oxide, just as we found in Section III for the 
prefactors of the dielectric constant: 
$B_{2LS} \simeq B_{bulk}(1-x)+B_{surf}x^f$, $B_{ATS}\simeq Bx$. 
Also $B_{ph}$ requires to be re-evaluated. With increasing K$_2$O molar 
concentration $x$ for the (SiO$_2$)$_{1-x}$(K$_2$O)$_x$ glass, the number of 
phonons from the NM-component (K$_2$O in this case) increase linearly with 
the concentration $x$, and for the NF-component (SiO$_2$) it should also
decrease linearly, like ($1-x$). Just as we assumed in the previous Section 
III, there are fractal/percolation effects between the NM- and NF-clusters, 
which makes room for some percolation clusters' interfaces where the phonons 
also might contribute somehow with a term proportional to $C_{ph}x^f$
($C_{ph}$ being an $x$-independent constant).

For these glasses, moreover, a non-negligible concentration of Fe$^{3+}$ (or,
according to coloring, Fe$^{2+}$) impurities is reported, a side effect of 
the industrial production process. Estimates give 102 ppm for AlBaSiO and 
126 ppm for Duran, 6 ppm for BK7, 100 ppm for Pyrex 7740 and 12 ppm for Pyrex 
9700 (see, e.g., the discussion in Ref \cite{Ref14}). All glasses may, indeed, 
contain some [FeO$_4$]$^{0}$ impurity-substitution F-centers (in the glass, 
similar to a liquid, in concentrations however much, much lower than the 
nominal Fe bulk concentrations \cite{Ref32}). The Fe$^{3+}$ cation and the 
O$^{2-}$ anion, on which the hole is localized (forming the O$^{-}$ species, 
that is the O$^{2-}$ 
$+$ hole subsystem), form a bound small polaron. In this configuration the 
Fe$^{3+}$ cation is subject to a crystal field with an approximate $C_{3}$ 
symmetry axis along the Fe$^{3+}$ - O$^{-}$ direction. This axis plays a 
quantization role for the Fe$^{3+}$ electronic spin. The hole is assumed to 
be tunneling between two neighboring oxygen ions, switching the quantization 
axis between two directions and therefore entangling its spin states. This 
is likely to give some tiny contribution to the heat capacity and we should, 
therefore, also take it here into account \cite{Ref44}. The spin Hamiltonian 
of the [FeO$_{4}$]$^{0}$ F-center is $H_{s-S}=V_zs_zS_z$, where $V_{z}$ is 
the principal value of the dipole interaction matrix, $s_z$ and $S_z$ are the 
spin operators of the hole and of the Fe$^{3+}$ ion, respectively. In the 
absence of a magnetic field there are only two low-lying energy levels: 
$E_{1,2}=\pm \frac{5}{4}\left|V_z\right|$. The unknown distribution function 
$G(V_{z})$ must approach zero when its argument approaches either zero or 
infinity and have a maximum at a definite argument value $V_{0}$. The simplest 
one-parameter function displaying such properties is a Poisson distribution:

\begin{equation}
G\left(V_z\right)=\frac{4V^2_z}{V^3_0}\exp\left(-\frac{2V_z}{V_0}\right), 
\quad V_z\in \left(-\infty;0\right], \quad V_0<0.
 \label{rtGVz}
\end{equation}
The contribution from the [FeO$_{4}$]$^{0}$ ensemble to the heat capacity is, 
as usual:

\begin{equation}
C_{Fe^{3+}}(T)=-T\frac{{\partial}^2F_{Fe^{3+}}}{\partial T^2},
 \label{rtCFe}
\end{equation}
where $F_{Fe^{3+}}(T)$ is the free energy of the [FeO$_{4}$]$^{0}$ ensemble, 
that one evaluates as:

\begin{eqnarray}
&& F_{Fe^{3+}}=-k_BT\ln\left(e^{-{E_1}/{k_BT}}+e^{-{E_2}/{k_BT}}\right)=
\nonumber \\
&& =F_{Fe^{3+}}=-k_BT\ln\left(2{\cosh\left(\frac{E}{2k_BT}\right)}\right),
\label{rtFFe}
\end{eqnarray}
(here $E=\frac{5}{4}\left|V_z\right|$). Using said distribution function for 
$G(V_{z})$, Eq. (\ref{rtGVz}), as well as the expression for 
$C_{Fe^{3+}}\left(T\right)$ from $F_{Fe^{3+}}\left(T\right)$, one can obtain 
an expression for the heat capacity from the trace [FeO$_{4}$]$^{0}$ centres
in the glass, and which should be added to the total heat capacity $C_{p}$:

\begin{eqnarray}
&& C^{Fe^{3+}}_p(T)=
\rho(x)n_jk_B\times
 \nonumber \\
&& \times\int^{\infty}_0{dV_z{\left(\frac{E}{2k_BT}\right)}^2
\cosh^{-2}\left(\frac{E}{2k_BT}\right)G(V_z)}
\nonumber \\  
&& =\rho(x)n_jk_B\int^{\infty}_0{dV_z\frac{25V^4_z}{16T^2}
\frac{1}{V^3_0}e^{(-\frac{2V_z}{V_0})}
\cosh^{-2}\left(\frac{5V_z}{8k_BT}\right)} , 
\nonumber \\ 
\label{rtCpFe} 
\end{eqnarray}
where $n_j=x_j/\rho(x)$ is the mass concentration of the tiny amount of
Fe$^{3+}$ ions (a very small fraction of the total bulk Fe-concentration) 
substituting the Si$^{4+}$ in the network.

Hence, the total heat capacity will be the sum of all these contributions 
(Eqs. (\ref{rthcapac}), (\ref{rtCAtsfinal}) and (\ref{rtCpFe})) \cite{Ref45}:

\begin{equation}
C_p(T)=B_{ph}T^3+B_{2LS}T+C^{ATS}_p(T)+C^{Fe^{3+}}_p(T).
 \label{rtCpSum}
\end{equation}
Making use of expression (\ref{rtCpSum}), we have fitted the experimental data 
for the heat capacity of the (SiO$_2$)$_{1-x}$(K$_2$O)$_x$ glasses from 
\cite{Ref18}. In order to fit the pure a-SiO$_{2}$ data we use only formula 
(\ref{rthcapac}), that fits the pure silica's data well within the 2LS-STM.

The heat capacity $C_{p}(T,x)$ data \cite{Ref18} for the 
(SiO$_2$)$_{1-x}$(K$_2$O)$_x$ glasses were obtained using a signal-averaging 
technique and for these samples the data are presented in Fig. \ref{fig5}. 
As one can see, the heat capacity for the 
(SiO$_2$)$_{1-x}$(K$_2$O)$_x$ glasses at low temperatures is larger than that 
for pure silica glass, as is typical for the multicomponent glasses, already 
with the smallest 5$\%$ concentration of K$_{2}$O. The heat capacity decreases 
and then again increases with increasing molar concentration $x$ of K$_{2}$O. 
The additional heat capacity arises from the addition of ATS in the K$_{2}$O
NM-clusters and also from the presence of Fe$^{3+}$ impurities, contained in 
small (and unknown) concentrations, but contributing to the low- and 
middle-range of the temperature dependence.

\begin{table}[h]
\begin{tabular}{|l|ccccc|}
\hline glass & $x$ & $B_{ph}\times10^8$ & $B_{2LS}\times10^8$ & $B_{ATS}
\times10^8$  & $x_j$ \\
type & mol & Jg$^{-1}$K$^{-4}$ & Jg$^{-1}$K$^{-2}$ & Jg$^{-1}$K$^{-1}$ & ppm \\
\hline \hline
SiO$_2$ & 0 & 245.55 & 70.65 & -  & -   \\
\hline
K-Si & 0.05 & 260.92 & 155.23 & 22.77 & 29.86  \\
\hline
K-Si & 0.08 & 266.36 & 196.11 & 36.44 & 18.15  \\
\hline
K-Si & 0.10 & 269.46 & 221.62 & 45.55 & 10.54  \\
\hline
K-Si & 0.20 & 281.42 & 337.19 & 91.11 & 3.00  \\
\hline
\end{tabular}
\caption[1]{ Extracted parameters for fits to the heat capacity data for 
SiO$_2$ and (SiO$_2$)$_{1-x}$(K$_2$O)$_x$ glasses, with $D_{min}=$0.87 K 
and $V_{0}=$-0.42 K as fixed.}
\label{tabl2}
\end{table}

\begin{figure}[h]
\includegraphics[width=0.48\textwidth]{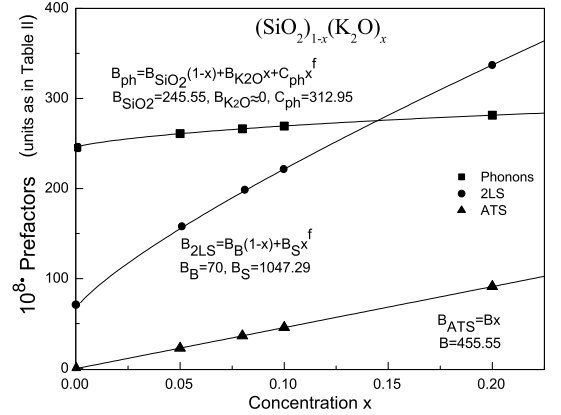} \vskip-5mm
\caption[2]{ The 2LS and ATS prefactor parameters $\left(\times {10}^8\right)$ 
for all glasses (from Table \ref{tabl2}) as a function of $x$. The 
experimental data fit well with our theoretical expectations with $f=0.81$ 
(full lines).  }
\label{fig6}
\end{figure}

Both prefactors, for 2LS and ATS, are indeed dependent on the molar 
concentration $x$ from our data analysis, and in the same way as we did in 
Section III we have fitted the extracted prefactors with the forms: 
$B_{2LS}\simeq B_{bulk}(1-x)+B_{surf}x^f$, $B_{ATS}\simeq Bx$ ($B$ being
some constant). These dependencies are show in Fig. \ref{fig6}. Also 
$B_{ph}$ is found to change by increasing the concentration $x$ of the good 
crystal-former, K$_2$O, and in the way we anticipated.

\noindent
With increasing concentration $x$, for the (SiO$_2$)$_{1-x}$(K$_2$O)$_x$ glass 
the number of phonons from the NM-component (K$_{2}$O) increases linearly 
with the concentration, and for the NF-component (SiO$_{2}$) it should be 
decreasing linearly like $(1-x)$. As we reasoned for the dielectric constant, 
there are percolation mixing effects between the NM- and the NF-systems, which 
create percolation clusters and their NF/NM interfaces where phonons also 
might be populated in a way proportional to $C_{ph}x^f$. As it turns out,
the very same value $f=0.81$ can be extracted from all our fits, just as was 
done in Section III for the dielectric constant data.

\section{V. SUMMARY AND CONCLUSIONS}

Thus, we have shown that there is direct evidence in zero magnetic field 
already for the existence of multiwelled ATS (modelled as tunneling 3LS) and 
with the new distribution function advocated to explain the magnetic-field 
effects in the multicomponent glasses (see \cite{Ref14}). The relevance of 
near-degenerate multiwelled TS in glasses is a new and unexpected finding in 
this field of research. Our work predicts, in particular, that the magnetic 
response of the mixed alkali-silicate glasses should be important and scale 
like the molar alkali concentration $x$. At the same time the -1:1 slope-ratio 
problem of the standard TM in comparison with experimental data for 
$\epsilon'(T)$ has been given a simple explanation in terms of our two-species 
tunneling model. The main result of this work is that the concentration 
$x_{ATS}$ (absorbed in $\tilde{P}^*$ and thus in the $A_{ATS}$- and 
$B_{ATS}$-prefactors) of ATS indeed scales linearly with $x$ for both 
$\epsilon'(T,x)$ and $C_p(T,x)$. This is supported by our analysis of the 
experimental data of Ref. \cite{Ref18} very well indeed. This analysis is, in 
our view, strong evidence that the ATS are nesting in the NM-clusters of the 
good crystal-formers. 

Using the results of this analysis (and for AlBaSiO the results of the 
experimental-data analysis in a magnetic field \cite{Ref14}) we can estimate 
the value of the dipole moment associated with the ATS, 
$p_{eff}=\sqrt{\overline{p_1^2}}$. For AlBaSiO, using the value of 
$\tilde{P}^{*}$ extracted from $C_p$ \cite{Ref14} and that of $A_{ATS}$ given 
in Table \ref{tabl1} we extract $p_{eff}$=0.41 D. For 
(SiO$_2$)$_{1-x}$(K$_2$O)$_x$, we notice from the definitions in Section 
III that the ratio of the dielectric and heat capacity prefactors,

\begin{equation}
\frac{A_{ATS}}{B_{ATS}}=
\frac{\rho(x)}{{2\epsilon_{0}\epsilon_{r}k_BD_{min}}\bar{{p_1}^2}}
 \label{rtRatioPref}
\end{equation}
is almost independent of the K$_2$O concentration $x$. From our extracted 
values in Tables \ref{tabl1} and \ref{tabl2} and the measured values of 
$\rho(x)$ \cite{Ref18} we estimate $p_{eff}=0.045$ D for the mixed glasses, 
independently of $x$! Considering the elementary atomic electric-dipole's 
value is $ea_0=2.54$ D, these small values of $p_{eff}$ for the ATS confirm 
that their physics must come from the coherent (or correlated) tunneling of 
small ionic clusters (the very same origin for the large values of $D_{min}$ 
and for $D_{0min,max}$, see the Appendix). Indeed, a cluster of $N$ 
coherently-tunneling particles has a dipole moment 
$p_{eff}=\left|{\sum}_{i=1}^N{\bf p}_i\right|$ that can become much smaller 
than $ea_0$ (the order of magnitude of each $\left| {\bf p}_i \right|$ in the 
sum) as $N$ grows large. The fact, that we extract values of $p_{eff}$ much 
smaller than $ea_0$, confirms the picture of a correlated tunneling cluster
in the $B=0$ case already.

It is noteworthy that several papers from the Anderson group have proved that 
the addition of any NM-species in a networking pure glass causes significant 
(and thus far unexplained) deviations from the predictions of the 2LS-STM
\cite{Ref18,Ref48}. We have explained the origin of these deviations, for 
$C_p(T,x)$ as well as for $\epsilon(T,x)$. However, experiments do show that 
the thermal conductivity $\kappa(T,x)\propto T^2$ remains (below 1 K) 
remarkably universal and composition-independent \cite{Ref18}. This is 
connected with the super-universality of the internal friction coefficient, 
$Q^{-1}$, in the cold glasses; these and other remarkable findings will be 
addressed elsewhere within the context of our approach.  

In summary, we have shown that there is direct evidence in zero magnetic 
field already for the multiwelled ATS advocated to explain the magnetic 
field effects in the multicomponent glasses. Similar $x$-dependent phenomena 
are to be expected for the low-$T$ anomalies of the MAS-type ceramic-glass of 
composition (SiO$_2$)$_{1-x}$(MgO)$_x$, which should also respond to the 
magnetic field \cite{Ref46} (just like the mixed alkali-silicates of this 
work should). One may remark, at this point, that any extension of the 2LS-STM 
enlarging the adjustable-parameter space is bound to improve agreement with 
the experimental data. In this paper we have shown that it was not just a 
matter of quantitative agreement, but qualitative as well. Whilst agreeing 
that the TM remains unsatisfactory, we stress that it is the only approach we 
know of that is versatile enough to allow for an interesting explanation of 
rather puzzling phenomena at low-$T$ in the real glasses. Furthermore, our 
two-species, multilevel TS model has been able to consistently explain a good 
number of different experimental data \cite{Ref14,Ref42}. It cannot be a mere 
coincidence that the same phenomenological model, with rather similar material 
parameters in different experiments, is capable of explaining so much new 
physics. Far from being an ad-hoc model, our approach reveals the intimate 
microscopic structure of the real glasses, which cannot be considered as 
being homogeneously-disordered anymore, and this must have some important 
consequences also for a better understanding of the mechanisms underlying 
the glass transition.         

As for the possibility of estimating the size and density of the incipient
crystals in glasses from our theory, we remark that the simplified 
geometric-averaging procedure adopted for the physics of the ATS so far 
\cite{Ref14} does not allow anything more than an estimate of the $P^*$ 
parameter ($\sim$ 1.97 $\times$ 10$^{17}$ cm$^{-3}$ for AlBaSiO \cite{Ref14}, 
this being in fact the value of $x_{ATS}P^*$, $P^*$ being the unknown 
dimensionless parameter of the ATS distribution in Eq. (\ref{rtProb3LS})). 
However, the geometric-averaging procedure should be performed in two stages 
(within the incipient micro-crystals first and then within the glassy matrix
in which the crystallites are embedded) at the price of making the theory 
considerably more complicated. When this is done, with a more efficient and 
complete theoretical formulation, then information on the size distribution 
of the incipient crystallites could be gained from further low-$T$ experiments 
in magnetic fields and at different controlled compositions.

\section{ACKNOWLEDGEMENTS}

Special thanks are due to M.F. Thorpe for advice, to A. Bakai, M.I. Dyakonov, 
P. Fulde and S. Shenoy for their interest in this work and for illuminating 
discussions, and to A. Borisenko for a fruitful collaboration. M.P. is 
grateful to the Italian Ministry of Education, University and Research (MIUR) 
for support through a Ph.D. grant of the ``Progetto Giovani (ambito di 
indagine n. 7: materiali avanzati (in particolare ceramici) per applicazioni 
strutturali).''

\section{APPENDIX}

\begin{figure}[h]
\includegraphics[width=0.48\textwidth]{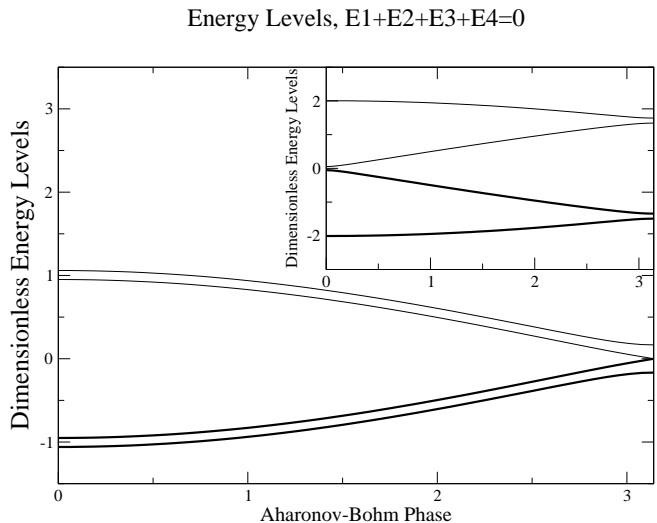} \vskip+2mm
\caption[2]{ (Main) Variation with the magnetic Aharonov-Bohm phase $\phi$
of the energy spectrum (units $D_2=1$) for the case $D_1=0$ and a choice of
$E_1, E_2, E_3, E_4$ with $\sqrt{E_1^2+E_2^2+E_3^2+E_4^2}/D_2=0.01$. This is
to be compared with the 3LS energy spectrum, Fig. \ref{3lsspectrum}. (Inset) 
The energy spectrum in the opposite case, $D_2=0$, and a choice of
$E_1, E_2, E_3, E_4$ with $D_1$ about 100 times stronger.}
\label{4lsspectrum}
\end{figure}

We first show how the spectrum of the 4LS Hamiltonian, Eq. (\ref{rtH4matrix}),
is similar to that of the 3LS Hamiltonian, Eq. (\ref{rtH3matrix}) in the
near-degenerate limit. We rewrite the $H_0^{(4)}$ 4LS Hamiltonian in the 
presence of a magnetic field, coupled orbitally to the charged tunneling 
particle:

\begin{equation}H_0^{(4)}=\left(\begin{array}{cccc}
E_1 & D_1e^{i\phi/4} & D_2e^{i\phi/2} & D_1e^{-i\phi/4} \\ 
D_1e^{-i\phi/4} & E_2 & D_1e^{i\phi/4} & D_2e^{i\phi/4} \\ 
D_2e^{-i\phi/4} & D_1e^{-i\phi/4} & E_3 & D_1e^{i\phi/4} \\
D_1e^{i\phi/4} & D_2e^{-i\phi/2} & D_1e^{-i\phi/4} & E_4
\end{array}\right)
\label{rtH4matrixB} 
\end{equation}
where $\sum_{i=1}^4E_i=0$ is imposed, $D_1$ and $D_2$ are the n.n. and n.n.n. 
hopping energies, respectively, and where

\begin{equation} 
\phi=2\pi\frac{\Phi(\bf B)}{\Phi_0}, 
\quad \Phi({\bf B})={\bf S}_{\diamondsuit}\cdot{\bf B}, \quad 
\Phi_0=\frac{hc}{q}
\end{equation}
is the Aharonov-Bohm phase resulting from the magnetic flux $\Phi({\bf B})$
threading the square-loop (having area $S_{\diamondsuit}$) closed trajectory 
of the particle. The above Hamiltonian should in fact be symmetrised over its 
permutations, since the sign of the n.n.n. Peierls phase is ambiguous (in 
practice, one replaces $D_2e^{\pm i\phi/2}$ with $D_2\cos(\phi/2)$ in the 
appropriate matrix entries). The eigenvalues equation giving the energy 
levels is then:

\begin{eqnarray}
&& {\cal E}^4
+{\cal E}^2 \Big( \sum_{i<j}E_iE_j-4D_1^2-D_2^2(1+\cos\phi) \Big) 
\nonumber \\
&& -{\cal E} \Big( \sum_{i<j<k}E_iE_jE_k+4D_1^2D_2(1+\cos\phi) \Big) \\
&& +E_1E_2E_3E_4-D_1^2 ( E_1E_2+E_2E_3+E_3E_4+E_4E_1 ) \nonumber \\
&& -\frac{1}{2}D_2^2 ( E_1E_3+E_2E_4 )(1+\cos\phi)-2D_1^2D_2^2(1+\cos\phi) 
\nonumber \\
&& +2D_1^4 ( 1-\cos\phi )+\frac{1}{8}D_2^4
\big( 3+4\cos\phi+\cos(2\phi) \big)=0. \nonumber
\label{eigeneq4} 
\end{eqnarray}
More instructive than numerically extracting the four exact roots 
${\cal E}_{0,1,2,3}$ (with ${\cal E}_0<{\cal E}_1<{\cal E}_2<{\cal E}_3$) is 
for us the physically interesting limit case in which $|E_i/D_1|\ll 1$ 
and $|D_2/D_1|\ll 1$ (near-degeneracy of the four-welled potential). The 
above eigenvalue equation then becomes much easier to study:

\begin{eqnarray}
&& {\cal E}^4
+{\cal E}^2 \Big( \sum_{i<j}E_iE_j-4D_1^2 \Big) \nonumber \\
&& -D_1^2 ( E_1E_2+E_2E_3+E_3E_4+E_1E_4 ) \nonumber \\
&& +2D_1^4 ( 1-\cos\phi ) \approx 0 
\end{eqnarray} 
this being the eigenvalue equation of the reduced 4LS Hamiltonian

\begin{equation}H_{0red}^{(4)}=\left(\begin{array}{cccc}
E_1 & D_1e^{i\phi/4} & 0 & D_1e^{-i\phi/4} \\ 
D_1e^{-i\phi/4} & E_2 & D_1e^{i\phi/4} & 0 \\ 
0 & D_1e^{-i\phi/4} & E_3 & D_1e^{i\phi/4} \\
D_1e^{i\phi/4} & 0 & D_1e^{-i\phi/4} & E_4
\end{array}\right)
\label{redtH4matrixB} 
\end{equation}
always for $|E_i/D_1|\ll 1$, and which has the following solutions:

\begin{eqnarray}
&& \frac{{\cal E}_{0,1,2,3}}{D_1}=\pm \frac{1}{\sqrt{2}} 
\Bigg\{ 4-\sum_{i<j}\frac{E_iE_j}{D_1^2}
\pm \Big[ \Big( 4-\sum_{i<j}\frac{E_iE_j}{D_1^2} \Big)^2 \nonumber \\
&& +4\frac{E_1E_2+E_2E_3+E_3E_4+E_4E_1}{D_1^2} \nonumber \\
&& + 8 ( \cos\phi-1 ) \Big]^{1/2} \Bigg\}^{1/2}.
\end{eqnarray}
The perhaps surprising result is an energy spectrum where only the middle
doublet (${\cal E}_1$, ${\cal E}_2$ in our notation) becomes near-degenerate
at weak (or zero) magnetic fields ($\phi\to 0$). This is shown in the inset
of Fig. \ref{4lsspectrum} and is reminiscent of the situation with dimerized 
2LS considered in Ref. \cite{Ref49} in order to account for the oscillations 
of the dielectric constant with $B$. Beside there being no evidence for a 
dimerization of TS in glasses (unlike perhaps in mixed and disordered 
crystals), one would have to explain why the ground state ${\cal E}_0$ is
prohibited for the tunneling particle (the real energy gap being in fact 
$\Delta{\cal E}={\cal E}_1-{\cal E}_0$). The way out can be found again in 
Sussmann's paper \cite{Ref34} since the $n_w=4$ welled trapping potential
giving rise to the same physics as our 3LS must in fact have tetrahedral 
and not square geometry. The tetrahedral 4LS in a magnetic field will be 
considered elsewhere \cite{Ref42}. Here we only want to remark that the 
tetrahedral situation can be mimicked by a square multiwelled potential in 
which $|D_1/D_2|\ll 1$ and always in the limit case $|E_i/D_2|\ll 1$. This 
corresponds to the counter-intuitive situation in which it is easier for the 
particle to tunnel across the square to the n.n.n. site rather than to a n.n. 
site, as if the middle potential barrier had collapsed. In this limit case,
Eq. (\ref{eigeneq4}) becomes, instead:

\begin{eqnarray}
&& {\cal E}^4
+{\cal E}^2 \Big( \sum_{i<j}E_iE_j-2D_2^2\cos ^2\frac{\phi}{2} \Big) 
\nonumber \\
&& -D_2^2 ( E_1E_3+E_2E_4 )+D_2^4\cos ^4\frac{\phi}{2} \approx 0 
\end{eqnarray}
with, once more, easily found solutions

\begin{eqnarray}
&& \frac{{\cal E}_{0,1,2,3}}{D_2}=\pm \frac{1}{\sqrt{2}} 
\Bigg\{ 2\cos ^2\frac{\phi}{2}-\sum_{i<j}\frac{E_iE_j}{D_2^2} \nonumber \\
&& \pm \Big[ \Big( 2\cos ^2\frac{\phi}{2}-\sum_{i<j}\frac{E_iE_j}{D_2^2} 
\Big)^2 \nonumber \\
&& +4\frac{E_1E_3+E_2E_4}{D_2^2} 
-4 \cos ^4\frac{\phi}{2} \Big]^{1/2} \Bigg\}^{1/2},
\end{eqnarray}
as exemplified in Fig. \ref{4lsspectrum} (main). We therefore obtain that
the lowest-lying gap remains near-degenerate for weak fields and we can 
conclude, therefore, that the lowest-lying eigenvalues display, at low-$T$, 
almost the same physics as in the case of a 3LS (Fig. \ref{3lsspectrum}). 
This shows the important role of the frozen solid surrounding the tunneling
``particle'' (which could be, perhaps, a vacuum in fact) and that when 
nested within an incipient crystallite a magnetic-field sensitive TS is well 
described by a tunneling 3LS as the minimal generic model potential. 

Next, we seek a description of a cluster of $N$ correlated tunneling 
particles (atoms, ions or molecules) and derive the transformation rules for 
the tunneling parameters (including also those involved in the theory for 
the magnetic effects \cite{Ref14}) when the cluster is replaced by a single 
``tunneling particle'' as the result of the coherent tunneling (CT) of the 
particles in the cluster.

The ``tunneling particle'' in question is only a fictitious one, as was 
inferred in Section II by examining local minima in the energy landscape, 
representing the CT of a cluster of $N$ true tunneling particles (which in 
the real glasses might be the lighter species involved in the material: 
Li$^+$ in the disordered crystal Li:KCl, 
O$^{2-}$ in the multisilicates and H$^+$ and/or D$^+$ in a-glycerol) and for 
which we have to make up appropriate renormalized tunneling parameters. The 
concept of CT in separate local potentials is distinct from that of the joint 
tunneling of $N$ particles in the same local potential, for in the latter case 
the tunneling probability would be depressed exponentially: 
$D_0/\hbar\approx\Omega(\frac{\Delta_0}{\hbar\Omega})^{\sqrt{N}}$ ($\Delta_0$ 
being the real particles' common tunneling transparency). As we shall show 
below, at least for moderate values of $N$, for CT in separate potentials 
we expect instead:

\begin{equation}
D_0\approx N\Delta_0, \qquad D_{min}\approx N\Delta_{min}
\label{renorm1}
\end{equation}
and, for the fictitious particle's charge and flux-threaded area (see 
\cite{Ref14} for the magnetic effects):

\begin{equation}
q=Nq_0, \qquad S_{\triangle}\approx 4Na_0^2
\label{renorm2}
\end{equation}
($q_0=O(e)$ being the charge of the real tunneling particles, $a_0$ Bohr's 
radius). In the latter relations, less obvious is the renormalization of the 
flux-threaded area $S_{\triangle}$ of a 3LS ATS. It is however the direct 
consequence of our multiphase model of a real glass, thought of as made up 
of regions of enhanced atomic ordering (RER) or micro-crystals (Figs. 
\ref{cembryo}, \ref{cartoon}) embedded in a homogeneously-disordered host 
matrix. The magnetic flux appears quadratically in our theory \cite{Ref14}, 
each elementary flux adding up within each micro-crystallite or RER and then 
appearing, squared, multiplied by $\cos^2\beta$ in the glassy matrix in a 
magnetic field ($\beta$ being the random angle formed by ${\bf S}_{\triangle}$ 
with the magnetic field ${\bf B}$), a factor averaging out to $\frac{1}{2}$ 
in the bulk. From these considerations and from Eqs. (\ref{renorm1}) and 
(\ref{renorm2}), the renormalization of the composite phenomenological 
parameter $D_0\vert\frac{q}{e}\vert S_{\triangle}$ would be as follows (if
$q=2e$, appropriate for the multisilicates):

\begin{equation}
D_0\vert\frac{q}{e}\vert S_{\triangle}\approx 8N^3\Delta_0a_0^2
\label{renorm3}
\end{equation}
Setting $\Delta_{0}$=1 mK, one gets a value of $N$ ranging from about 25 
coherent-tunneling particles in a cluster at the lowest temperatures 
\cite{Ref42}, to about 600 at the higher temperatures. These estimates are 
somewhat speculative, since the real values of the elementary flux-threaded 
area and of the elementary tunneling barrier transparency $\Delta_0$ are 
unknown, we are however inclined to support the value $N\approx$ 200 that was 
proposed by Lubchenko and Wolynes \cite{Ref22}. This would yield a value of 
$\Delta_{min}$ ranging from 80 $\mu$K to 4 mK also for the mixed 
alkali-silicate glasses (for which $D_{min}\approx 800$ mK). The above 
considerations show all in all the tendency for the coherent-tunneling 
cluster size $N$ to be also temperature dependent. 

We now come to the justification of Eq. (\ref{renorm1}). At low temperatures 
the interactions between true tunneling particles become important and 
coherent-tunneling motion can take place. Coherent motion in the context of 
the tunneling model is a state in which all of the particles in each local 
potential contribute to the overall tunneling process in a correlated way. 
We exemplify our ideas in the context of the simplest 2LS situation first.
Let us consider two interacting 2LS. Let the positions of the particles in 
the two wells be left (L) and right (R). The tunneling particles in the 
cluster interact via a weak potential $U$ which may have its origin, for
example, from either a strain-strain interaction having the form
$U\sim A/r^3$ (dipole-dipole interaction) \cite{Ref22,Ref47}, where $r$ is 
the distance between a pair of tunneling particles either in the L or R 
wells and $A$ is a constant, or it could be due to electrostatic dipole-dipole 
interaction. The tunneling of the particle in one 2LS from L to R (or vice 
versa) influences, via the interaction, the particle in the other 2LS, forcing 
it to jump into the free well. The hopping Hamiltonian of two interacting 2LS 
can be written as follow (with $\Delta_{iL}=-\Delta_{iR}=\Delta_i$ and 
dropping factors of -1/2):

\begin{eqnarray}
H_2 &=& \sum_{a=L,R}\Delta_{1a}c^{\dagger}_{1a}c_{1a}+ 
\Delta_{01}\sum_{a\not=a'}c^{\dagger}_{1a}c_{1a'}+{\rm hc} \nonumber \\
&+& \sum_a\Delta_{2a}c^{\dagger}_{2a}c_{2a}+ 
\Delta_{02}\sum_{a\not=a'}c^{\dagger}_{2a}c_{2a'}+{\rm hc} \nonumber \\
&-& U\left ( c^{\dagger}_{1L}c_{1L}c^{\dagger}_{2L}c_{2L}
+c^{\dagger}_{1R}c_{1R}c^{\dagger}_{2R}c_{2R} \right )
\label{int2H2}
\end{eqnarray} 
which favors coherent LL$\to$RR and RR$\to$LL joint tunneling and acts on 
the joint states 
$\vert aa' \rangle=\vert LL \rangle,\vert LR \rangle,\vert RL \rangle,
\vert RR \rangle$. The coherent motion of the two real particles can now be 
replaced by the tunneling of a new, fictitious particle in its own double 
well. In order to write the renormalized Hamiltonian of two coherent- 
tunneling particles we are interested only in the matrix elements
$\langle LL \vert H_2 \vert LL \rangle$, 
$\langle RR \vert H_2 \vert RR \rangle$,
$\langle RR \vert H_2 \vert LL \rangle$ and
$\langle LL \vert H_2 \vert RR \rangle$ of Hamiltonian (\ref{int2H2}):

\begin{eqnarray}
\langle LL \vert H_2 \vert LL \rangle &=& \Delta_1+\Delta_2-U \\
\langle RR \vert H_2 \vert RR \rangle &=& -\Delta_1-\Delta_2-U \nonumber \\
\langle RR \vert H_2 \vert LL \rangle &=& 
\langle LL \vert H_2 \vert RR \rangle=\Delta_{01}+\Delta_{02} \nonumber
\end{eqnarray}
(instead of the latter two, the pair $\langle RL \vert H_2 \vert LR \rangle$ 
and $\langle LR \vert H_2 \vert RL \rangle$, having the very same value 
$\Delta_{01}+\Delta_{02}$, could have served the purpose). These matrix 
elements represent the Hamiltonian of the fictitious particle, which 
corresponds to both real particles tunneling coherently together:

\begin{equation}
H_1'=\left(\begin{array}{cc}
\Delta_1+\Delta_2-U & \Delta_{01}+\Delta_{02} \\ 
\Delta_{01}+\Delta_{02}  & -\Delta_1-\Delta_2-U
\end{array}\right).
\label{renH2}  
\end{equation}
(the condition $\Delta'_1+\Delta'_2=0$ to be fixed through the addition of an
overall constant). Next, we consider the case of three interacting 2LS and 
repeat the previous considerations. The Hamiltonian of three interacting 2LS 
has the form:

\begin{eqnarray}
H_3 &=& \sum_{i=1}^3 \left \{ \sum_{a=L,R}\Delta_{ia}c^{\dagger}_{ia}c_{ia} 
+\sum_{a\not=a'}(\Delta_{0i}c^{\dagger}_{ia}c_{ia'}+{\rm hc}) \right \} 
\nonumber \\
&-& U\sum_{i<i'}\sum_a c^{\dagger}_{ia}c_{ia}c^{\dagger}_{i'a}c_{i'a} 
\label{int3H2}
\end{eqnarray} 
The matrix elements of a single replacing fictitious particle that correspond 
to CT are obtained as follows:

\begin{eqnarray}
\langle LLL \vert H_3 \vert LLL \rangle &=& \Delta_1+\Delta_2+\Delta_3-3U  \\
\langle RRR \vert H_3 \vert RRR \rangle &=& -\Delta_1-\Delta_2-\Delta_3-3U 
\nonumber \\
\langle RRR \vert H_3 \vert LLL \rangle &=& 
\langle LLL \vert H_3 \vert RRR \rangle=\Delta_{01}+\Delta_{02}+\Delta_{03}
\nonumber 
\end{eqnarray}
(the choice of the latter two not excluding the remaining coherent matrix 
elements pairs: 
$\langle RRL \vert H_3 \vert LLR \rangle$ and
$\langle LLR \vert H_3 \vert RRL \rangle$, 
$\langle RLR \vert H_3 \vert LRL \rangle$ and
$\langle LRL \vert H_3 \vert RLR \rangle$, 
$\langle LRR \vert H_3 \vert RLL \rangle$ and
$\langle RLL \vert H_3 \vert LRR \rangle$, which are all equivalent).
One can notice that the renormalized tunneling parameter is the sum of the 
$\Delta_{0i}$ of each 2LS. The energy asymmetry is also the arithmetic sum of 
the $\Delta_i$ of each 2LS, but one must add the interaction energy $-U$
multiplied by $N(N-1)/2$. Thus, for a coherently-tunneling cluster of $N$ 2LS 
we find that the diagonal matrix element becomes, generalizing to arbitrary 
$N$: $\Delta=\sum_{i=1}^N\Delta_i-\frac{N(N-1)}{2}U$ and the off-diagonal 
element, that corresponds to the CT-splitting for all $N$ particles, becomes 
simply $\Delta_0=\sum_i\Delta_{0i}$.

Applying the previous considerations to our model for a number $N$ of ATS 
with three wells (see Fig. \ref{tricone}) we can write the interacting 
Hamiltonian in the form:

\begin{eqnarray}
H_N &=& \sum_{i=1}^N \left \{ \sum_{a=1}^3 E_{ia}c^{\dagger}_{ia}c_{ia}
+\sum_{a\not=a'}(D_{0i}c^{\dagger}_{ia}c_{ia'}+{\rm hc}) \right \} \nonumber \\
&-& U\sum_{i<i'}\sum_a c^{\dagger}_{ia}c_{ia}c^{\dagger}_{i'a}c_{i'a}
\label{intNH3}
\end{eqnarray} 
If we represent the group of $N$ coherently-tunneling particles as a single 
fictitious particle moving in a three-welled potential, which is characterized 
by its own ground state energies $E_A$ and tunneling parameter $D_0$, we can 
describe this renormalized 3LS by the following Hamiltonian:

\begin{equation}
H_1'={\sum}_{A=1}^{3} E_A{c_A}^\dagger c_A + 
{\sum}_{A\not=A'}D_0{c_A}^\dagger c_{A'}+{\rm hc}
\label{renH3} 
\end{equation}
The ground states energies $E_A$ in the wells and tunneling parameter $D_0$ 
for the fictitious particle, in line with the calculations above, can be 
obtained through:

\begin{eqnarray}
E_A &=& \langle aa\dots a \vert H_N \vert aa\dots a \rangle \qquad A=a=1,2,3 
\nonumber \\
D_0 &=& \langle a'a'\dots a' \vert H_N \vert aa\dots a \rangle \\
&=& \langle aa\dots a \vert H_N \vert a'a'\dots a' \rangle \qquad a \not = a', 
\nonumber
\end{eqnarray}
(and the remaining variants of the second definition line, all equivalent). 
In analogy with the 2LS considerations, one can see that the renormalized
tunneling parameters $D=\sqrt{E_1^2+E_2^2+E_3^2}$ and especially $D_0$ can 
be replaced by the arithmetic sums of those of the bare coherently tunneling 
particles, $D\approx ND_i$ (neglecting the correction for a sufficiently 
weak $U$ and moderate values of $N$) and $D_0\approx ND_{0i}$, respectively. 
Indeed, the tunneling 
probabilities of weakly-correlated events should add up for values of $N$ not 
too large. Therefore, since $N$ can attain values as large as 200 \cite{Ref22} 
(independently of the solid's composition) in some models, and as corroborated 
by our reasoning in this Appendix, this leads to values of $D_i$ and $D_{0i}$ 
(as extracted from our theory's fitting parameters) comparable to those 
characteristic of the 2LS TM. The large values of $D_{min}$ and especially of 
$D_{0min}$ and $D_{0max}$, as extracted from our fits of our theory to the 
available experimental data, find therefore an interesting and physically 
plausible explanation. 

\begin{figure}[h]
\includegraphics[width=0.40\textwidth]{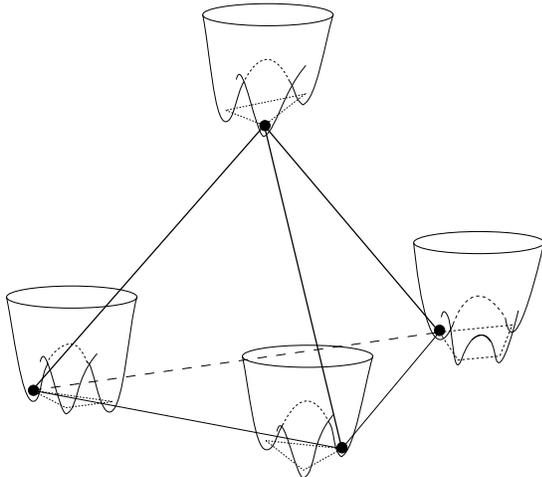} \vskip-5mm
\caption[2]{ A cluster of $N=4$ weakly interacting (real) tunneling 
particles that is being replaced with a (fictitious) single 3LS (Fig. 
\ref{tricone}(c)) having renormalised parameters according to Eqs.
(\ref{renorm1}) and (\ref{renorm2}). }
\label{cluster}
\end{figure}


\begin{thebibliography}{14}
\bibitem{Ref20} We remark that recently, thanks also to the efforts towards 
the explanation of the magnetic effects \cite{Ref14,Ref21} and from the 
study of quantum domain-wall excitations in the cold glasses \cite{Ref22}, a 
picture is emerging of a correlated (or coherent) tunneling cluster of some 
$N$ (charged) particles (atoms or molecules) which is being represented, in 
the TM, by a single fictitious tunneling particle. In this paper we argue 
$N\sim 200$ in agreement with Ref. \cite{Ref22}
\bibitem{Ref43d} We are well aware that the quoted fractal dimensions apply 
to percolation clusters, strictly, only at the percolation threshold $x=x_c$.
Our attitude is that fractal-type clusters can be used to model the NM-lumps
from the good crystal-formers even for $x<x_c$ and with non-integer fractal
dimensions. This assumption failing, we have no explanation for the extracted 
non-zero value of the $f$-exponent and $A_{surf}$-prefactor.
\bibitem{Ref45} Regarding the role of the Fe-impurities at zero magnetic 
field, this was totally overlooked in Ref. \cite{Ref14} where, to provide a
good fit of the data, the existence of a weak, stray magnetic field was 
wrongly advocated.
\bibitem{Ref46} Experiments on these glasses, using different isotopes 
Mg$^{24}$-Mg$^{26}$ and Mg$^{25}$, could also serve to confirm the nuclear 
quadrupole explanation for the magnetic effects.
\end{thebibliography}
\end{document}